\documentclass[a4paper]{article}
\usepackage{RR}
\RRNo{6510}
\usepackage{hyperref}
\usepackage{paralist}
\usepackage{graphicx}
\usepackage{amsmath}
\usepackage{wasysym}
\usepackage{amsfonts}
\usepackage{url}
\usepackage{algorithm}
\usepackage{algorithmic}
\usepackage{array}
\usepackage{multirow}
\newtheorem{definition}{Definition}[section] %
\newtheorem{example}{Example}[section] %
\newcommand{\Nat}{{\mathbb N}}
\usepackage[all,2cell,ps]{xy}

\RRdate{Avril 2008}
\RRauthor{% les auteurs
Hanifa Boucheneb\thanks{Laboratoire VeriForm, \'{E}cole Polytechnique de Montr\'{e}al, Canada (\texttt{hanifa.boucheneb@polymtl.ca}).}%
and
Abdessamad Imine\thanks{LORIA \& Univ. Nancy 2, UMR 7503 (\texttt{imine@loria.fr}).}%
}
\authorhead{H. Boucheneb and A. Imine}
\RRtitle{Exp\'eriences du Model-Checking des Algorithmes de R\'eplication Optimiste}
\RRetitle{Experiments in Model-Checking Optimistic Replication Algorithms} %%
\titlehead{Experiments in Model-Checking Optimistic Replication Algorithms}

\RRresume{Ce papier pr\'esente une s\'erie  d'exp\'eriences
pour v\'erifier des algorithmes de r\'eplication optimiste bas\'es sur
l'approche des transform\'ees op\'erationnelles qui est utilis\'ee
pour supporter l'\'edition collaborative.
Moyennant l'outil \emph {UPPAAL},
nous d\'ecrivons
formellement  le
comportement, la propri\'et\'e principale de consistance (\textit{i.e.} propri\'et\'e de convergence)
des syst\`emes d'\'edition collaborative,
ainsi qu'une abstraction du comportement de
l'environnement o\`u ces syst\`emes doivent op\'erer.
Ces syst\`emes comportent un nombre infini d'\'etats \`a cause de
leur caract\`ere interactif et de la
r\'eplication de donn\'ees.
Aussi, nous montrons comment exploiter les sp\'ecificit\'es de ces
syst\`emes ainsi que celles de l'outil \emph{UPPAAL}
pour att\'enuer la forte explosion d'\'etats de tels syst\`emes.

Deux mod\`eles, appel\'es respectivement mod\`ele concret et
mod\`ele symbolique sont propos\'es.\\ Le mod\`ele concret est
tr\`es proche de l'impl\'ementation du syst\`eme, mais se heurte
\`a une forte explosion d'\'etats. Pour att\'enuer cette explosion
d'\'etats, deux r\'eductions sont propos\'ees au mod\`ele concret.
La premi\`ere r\'eduction consiste \`a s\'electionner dans les
diff\'erents sites, d\`es le d\'ebut et de fa\c{c}on synchrone,
toutes les signatures des op\'erations \`a ex\'ecuter. La seconde
r\'eduction vise \`a synchroniser, certaines ex\'ecutions
d'op\'erations, si cela n'alt\`ere pas la propri\'et\'e de
convergence. Ces deux r\'eductions ont permis de r\'eduire, de
fa\c{c}on significative, la taille de l'espace d'\'etats mais ne
sont pas suffisantes pour v\'erifier certains algorithmes de
transformation consid\'er\'es ici.\\ Le mod\`ele symbolique vise
\`a pallier cette limitation en retardant la s\'election effective
et l'ex\'ecution des op\'erations jusqu'\`a la fin de la
construction des traces d'ex\'ecution de tous les sites (les
traces symboliques). Pour plus d'abstractions, ces \'etapes sont
encapsul\'ees dans une fonction et ex\'ecut\'ees de fa\c{c}on
atomique lors d'une transition d'\'etat. Cette fonction est
arr\^et\'ee aussit\^ot que la violation de la propri\'et\'e de
convergence est d\'etermin\'ee. Les deux r\'eductions propos\'ees
pour le mod\`ele concret sont aussi appliqu\'ees sur les
op\'erations symboliques.
 Les r\'esultats exp\'erimentaux ont montr\'e que le mod\`ele symbolique
 permet un gain significative en temps et en espace.
En utilisant le mod\`ele symbolique, nous avons pu montrer que si
le nombre de sites est  plus grand que $2$ alors la propri\'et\'e
de convergence n'est pas satisfaite pour
 tous les algorithmes de transformation consid\'er\'es. Un
 contrexemple est fourni pour chaque algorithme.}

\RRabstract{This paper describes a series of model-checking
  experiments to verify
optimistic replication algorithms based on Operational Transformation
(OT) approach used for supporting collaborative edition.
We formally define, using tool \emph{UPPAAL}, the behavior and the
main consistency requirement (\textit{i.e.} \emph{convergence property}) of the collaborative editing
systems, as well as the abstract behavior of
the environment where these systems are supposed to operate.
Due to data replication and the unpredictable nature of user
  interactions, such systems have infinitely many states. So,
we show how to exploit some features of
 the  \emph{UPPAAL} specification language to
attenuate the severe state explosion problem.
Two models are proposed. The first one, called \emph{concrete
model}, is very close to the system implementation but runs up
against a severe explosion of states. 
The second model, called \emph{symbolic model}, aims to overcome the
limitation of the concrete model by delaying the effective
selection and execution of editing operations until the
construction of symbolic execution traces of all sites is
completed. Experimental results have shown that the symbolic model
allows a significant gain in both space and time. 
Using the symbolic  model, we have been able to
show that if the number of sites exceeds $2$ then
the convergence property is not
satisfied for all OT algorithms considered here. A
counterexample is provided for every algorithm.} 
\RRmotcle{Algorithmes de transformation, Convergence des copies, Model-checking,v\'erification \`a la vol\'ee.}
\RRkeyword{Operational transformation algorithms, Copies convergence,
  Model-checking, on-the-fly approach.} 
\RRprojet{CASSIS}  % cas d'un seul projet
\RRtheme{\THSym} % cas d'un seul theme
\URLorraine % pour ceux qui sont \`a l'est
\begin{document}
\makeRR
   % cas d'un rapport de recherche
%% \makeRT % cas d'un rapport technique.
%% a partir d'ici, chacun fait comme il le souhaite
\tableofcontents

\section{Introduction}
This paper considers distributed collaborative editing systems. In such
systems, two or more users (sites) may manipulate simultaneously
some objects like texts, images, graphics etc. In order to achieve
an unconstrained group work, the shared objects are replicated at
the local memory of each participating user. Every operation is
executed locally first and then broadcast for execution at other
sites. So, the operations are applied in different orders at
different replicas of the object. This potentially leads to
divergent (or different) replicas, an undesirable situation for
replication-based distributed collaborative editing systems. \emph{Operational
Transformation} (OT) is an approach which has been proposed to overcome
the divergence problem~\cite{Ellis89}. In this approach, each non local operation
has to be transformed by applying some OT algorithm
before its execution. The main objective of this algorithm is to
ensure the \emph{convergence property}, \textit{i.e.} the fact that
all users view the same data.

In this work, we investigate use of a model-checking technique to verify
whether some OT algorithm satisfies the convergence
property or not. Model-checking is a very attractive and automatic
verification technique of systems. It is applied by representing
the behavior of a system as a finite \emph{state transition
system}, specifying properties of interest in a temporal logic
(\emph{LTL, CTL, CTL*, MITL, TCTL}) or a (timed)  B\"{u}chi
automaton and finally exploring the state transition system to
determine whether they hold or not. The main interesting feature
of this technique is the production of counterexamples in case of
unsatisfied properties. Several Model-checkers have been proposed
in the literature. The well known are
\emph{SPIN}\footnote{http://spinroot.com},
\emph{UPPAAL}\footnote{http://www.uppaal.com} and
\emph{NuSMV}\footnote{http://nusmv.irst.itc.it}.
Among these Model-checkers, we consider here the tool \emph{UPPAAL}.

\emph{UPPAAL} is a tool suite for validation and symbolic
model-checking of real-time systems. It consists of a number of
tools including a graphical editor for system descriptions (based
on Autograph), a graphical simulator, and a symbolic
model-checker. This choice is motivated by the interesting
features of \emph{UPPAAL} tools \cite{Larsen}, especially the
powerful of its description model, its simulator and its symbolic
model-checker. Indeed, its description model is a set of timed
automata \cite{Alur} extended with binary channels, broadcast
channels, C-like types, variables and functions. It is, at once,
simple and less restrictive comparing with description models of
other model-checkers. Its simulator is more useful and convivial
as it allows, in addition, to get and replay, step by step,
counterexamples obtained by its symbolic model-checker. Its
model-checker\footnote{The model-checker is used without the
graphical interface, i.e. tool \emph{memtime}}, based on a forward
on-the-fly method, allows to
compute over 5 millions of states.

To verify OT algorithms, we formally describe, using \emph{UPPAAL}, two
models and the requirements of the replication-based distributed collaborative
editing systems, as well as the abstract behavior of the
environment where these systems are supposed to operate. In the
first one, called \emph{concrete model}, the selection of operation
signatures and their effective execution are performed before or
during the generation of execution traces of different sites. To
attenuate the state explosion problem due to the different
interleaving of operations executed at different sites, we propose
to group, in one step, the execution of some of these operations
if this does not alter the convergence property.

In the second
model, called \emph{symbolic model}, the selection of operation
signatures and their effective execution are performed after
achieving the construction of execution traces of all sites
(symbolic traces). To make more abstractions, these steps are
encapsulated in a function executed as an atomic action. This
function is stopped as soon as the violation of the convergence
property is detected. Experimental results have shown that the
second model allows a significant gain in both space and time.
Another source
of the state explosion problem is the timestamp vectors used to
determine the dependency relation between operations. To attenuate
this state explosion, a variant of the symbolic model, where the
dependency relation is fixed and considered as an input data of
the symbolic model, is proposed.
Using the symbolic model, we have been able to
show that if the number of sites exceeds $2$ then
the convergence property is not
satisfied for all OT algorithms considered here. For every algorithm,
we provide a counterexample.

The paper starts with a presentation of the
OT approach and some of the known
OT algorithms proposed in the literature for synchronizing
shared text documents (Section $2$). Sections $3$
and $4$ are devoted to the description of both formal models and
their model-checking.
Related work and conclusion are presented respectively in
sections $5$ and $6$.

\section{Operational Transformation Approach}
\subsection{Background}
Operational Transformation (OT) is an optimistic replication  technique which allows many users
(or sites) to concurrently update the  shared data and next to
synchronize their divergent replicas in order to obtain the same
data. The updates of each  site are  executed on the  local
replica  immediately without being blocked or delayed, and then
are propagated to other sites to be executed again.  Accordingly,
every update is processed in four steps:
\begin{inparaenum}[(i)]
\item \textit{generation} on one site;
\item \textit{broadcast} to other sites;
\item \textit{reception} on one site;
\item \textit{execution} on one site.
\end{inparaenum}

%OT  considers  $n$   sites,  where  each  site  has   a  copy  of  the
%collaborative  object.
\medskip
\noindent\textbf{The shared object.} We deal with a shared object that admits a
linear structure.  To represent this  object we use  the \textit{list}
abstract data type.  A \textit{list}  is a finite sequence of elements
from a data type $\mathcal{E}$. This  data type is only a template and
can be instantiated by many  other types. For instance, an element may
be regarded as a character, a paragraph, a page, a slide, an XML node,
etc. Let $\mathcal{L}$ be the set of lists.

\medskip
\noindent\textbf{The primitive  operations.} It is assumed that  a list state
can only be modified by the following primitive operations:
\begin{itemize}
\item $Ins(p,e)$ which inserts the element $e$ at position $p$;
\item $Del(p)$ which deletes the element at position $p$.
\end{itemize}
%[[[\textbf{plus tu utilises c comme élément/caractère. Je ne sais
%pas s'il faut unifier }]]]

We assume that positions are given by natural numbers. The set of
operations is defined as follows:
\[
\mathcal{O} =\{Ins(p,e) | e\in
\mathcal{E}\mbox{ and } p\in\Nat\} \cup \{Del(p) | p\in\Nat\} \cup
\{Nop\}
\]
where $Nop$ is the idle operation that has null effect on the list state.
Since  the shared object is replicated, each site will own  a
local state  $l$ that  is altered  only  by local operations. The
initial state, denoted by $l_0$, is the same for all sites. The
function $Do   : \mathcal{O}\times\mathcal{L}\rightarrow
\mathcal{L}$, computes the state $Do(o,l)$ resulting from applying
operation $o$ to state $l$.   We   say  that   $o$  is
\textit{generated} on state $l$. We denote by
$[o_1;o_2;\ldots;o_n]$ an operation sequence. Applying  an
operation sequence to a  list $l$ is recursively defined as
follows:
\begin{inparaenum}[(i)]
\item $Do([],l)=l$, where $[]$ is the empty sequence and;
\item $Do([o_1;o_2;\ldots;o_n],l)=Do(o_n,Do(\ldots, Do(o_2, Do(o_1,l))))$.
\end{inparaenum}
Two operation  sequences $seq_1$ and  $seq_2$ are \textit{equivalent},
denoted by $seq_1\equiv  seq_2$,
iff  $Do(seq_1,l) =  Do(seq_2,l)$ for all lists $l$.

\begin{definition}\textbf{(Causality Relation)}\label{Def:caus}
Let an operation $o_1$ be generated at site $i$ and an  operation $o_2$
be generated at site $j$. We say that $o_2$ \emph{causally depends} on
$o_1$, denoted $o_1 \rightarrow o_2$, iff:
\begin{inparaenum}[(i)]
\item $i=j$ and $o_1$ was generated before $o_2$; or,
\item $i\neq j$ and the execution of $o_1$ at site $j$ has happened before
      the generation of $o_2$.
\end{inparaenum}
\end{definition}

\begin{definition}\textbf{(Concurrency Relation)}\label{Def:conc}
Two operations $o_1$ and $o_2$ are said to be \emph{concurrent},
denoted by $o_1 \parallel o_2$, iff neither $o_1 \rightarrow o_2$ nor
$o_2 \rightarrow o_1$.
\end{definition}

As a long established convention in OT-based distributed systems
(\textit{e.g.} collaborative editors)~\cite{Ellis89,Sun98}, the
\emph{timestamp vectors} are used to determine the causality and
concurrency relations between operations. Every timestamp is a vector
$V$ of integers with a number of entries equal to the number of sites.
For a site $j$, each entry $V[i]$ returns the number of operations
generated at site $i$ that have been already executed on site $j$.
When an operation $o$ is generated at site $i$, $V[i]$ is incremented
by $1$. A copy $V_o$ of $V$ is then associated to $o$ before its
broadcast to other sites. Once $o$ is received at site $j$, if the local
vector $V_{s_i}$ ``dominates''\footnote{We say that $V_1$ dominates $V_2$
  iff $\forall$ $i$, $V_1[i]\geq V_2[i]$.} $V_o$, then $o$ is ready
to be executed on site $j$. In this case, $V_{s_i}[i]$ will be
incremented by $1$ after the execution of $o$.
Otherwise, the $o$'s execution is delayed.

Let $o_1$ and $o_2$ be two operations issued respectively at sites
$s_{o_1}$ and $s_{o_2}$ and equipped with their respective timestamp
vectors $V_{o_1}$ and $V_{o_2}$. The causality and concurrency
relations are detected as follows:
\begin{itemize}
\item $o_1 \rightarrow o_2$ iff $V_{o_1}[s_{o_1}] > V_{o_2}[s_{o_1}]$;
\item $o_1 \parallel o_2$ iff $V_{o_1}[s_{o_1}] \leq V_{o_2}[s_{o_1}]$
  and $V_{o_1}[s_{o_2}] \geq V_{o_2}[s_{o_2}]$.
\end{itemize}

In the following, we define the conflict relation between two insert
operations:

\begin{definition}\textbf{(Conflict Relation)}\label{Def:conf}
  Two   insert  operations   $o_1   =  Ins(p_1,e_1)$   and  $o_2   =
  Ins(p_2,e_2)$,  generated on  different sites,  \emph{conflict} with
  each other iff:
\begin{inparaenum}[(i)]
\item $o_1 \parallel o_2$;
\item $o_1$ and $o_2$ are generated on the same list state; and,
\item $p_1 = p_2$, \textit{i.e.} they have the same insertion position.
\end{inparaenum}
\end{definition}

\textbf{To better understand our work, all examples given in this
  report use characters as elements to be inserted/deleted during a
  collaboration session.}

\subsection{Transformation principle}
A crucial issue  when designing  shared objects
with  a  replicated architecture  and  arbitrary messages communication
between sites  is the  \textit{consistency  maintenance} (or
\textit{convergence})  of all  replicas. To  illustrate  this problem,
consider the following example:

\begin{example}\label{exmp:e11}
  Consider   the   following   group   text   editor   scenario   (see
  Figure~\ref{fig:incons}): there  are two users (on two sites)  working on a
  shared  document represented  by  a sequence  of characters.   These
  characters  are addressed  from  $0$ to  the  end of  the document.
  Initially,  both copies hold  the string `` \emph{efecte}''.  User
  $1$  executes  operation  $o_1 =  Ins(1,\mbox{\emph{f}})$  to
  insert the character  \emph{f} at position $1$.  Concurrently,
  user  $2$   performs  $o_2  =  Del(5)$  to   delete  the  character
  \emph{e}  at  position  $5$.   When  $o_1$  is  received  and
  executed   on   site   $2$,   it  produces   the   expected   string
  ``\emph{effect}''.  But,  when $o_2$ is received on  site $1$, it
  does not take  into account that $o_1$ has  been executed before it
  and it produces the  string ``\emph{effece}''.  The result at site
  $1$  is different  from the  result of  site $2$  and  it apparently
  violates  the   intention  of   $o_2$  since  the   last  character
  \emph{e}, which  was intended to be deleted,  is still present
  in the  final string.   Consequently, we obtain  a \emph{divergence}
  between sites $1$ and $2$.  It  should be pointed out that even if a
  serialization protocol~\cite{Ellis89}  was used to  require that all
  sites execute $o_1$  and $o_2$ in the same  order (\textit{i.e.} a
  global order on concurrent operations) to obtain an identical result
  \emph{effece},  this identical  result  is still  inconsistent
  with the original intention of $o_2$.

\begin{figure}[t]%[htbp]
 \begin{minipage}[t]{0.5\linewidth}
\centerline{\xymatrix@C=10pt@M=2pt@R=10pt{
*+[F-,]\txt{site 1 \\ ``efecte''} \ar@{.}'[d]'[dd]'[ddd][dddd] &
*+[F-,]\txt{site 2 \\ ``efecte''} \ar@{.}'[d]'[dd]'[ddd][dddd] \\
o_1=Ins(1,f) \ar[ddr]  & o_2=Del(5) \ar[ddl] |!{[l];[dd]}\hole \\
*+[F]{\txt{``effecte''}} & *+[F]{\txt{``efect''}} \\
  Del(5)     &   Ins(1,f) \\
*+[F]{\txt{``effece''}}  & *+[F]{\txt{``effect''}} \\
}}
  \caption{Incorrect integration.}
  \label{fig:incons}
 \end{minipage}
 \begin{minipage}[t]{0.5\linewidth}
\centerline{\xymatrix@C=20pt@M=2pt@R=10pt{
*+[F-,]\txt{site 1 \\ ``efecte''} \ar@{.}'[d]'[dd]'[ddd][dddd] &
*+[F-,]\txt{site 2 \\ ``efecte''} \ar@{.}'[d]'[dd]'[ddd][dddd] \\
o_1=Ins(1,f) \ar[ddr]  & o_2=Del(5) \ar[ddl] |!{[l];[dd]}\hole \\
*+[F]{\txt{``effecte''}} & *+[F]{\txt{``efect''}} \\
IT(o_2,o_1)=Del(6)     & Ins(1,f) \\
*+[F]{\txt{``effect''}}  & *+[F]{\txt{``effect''}} \\}}
  \caption{Integration with transformation.}
  \label{fig:transf}
 \end{minipage}

\end{figure}

\end{example}

To maintain convergence, the OT approach has been proposed by
\cite{Ellis89}. When  User $X$  gets an operation $o$ that
was previously executed  by  User $Y$ on  his replica of the
shared  object   User $X$  does  not  necessarily  integrate $o$  by
executing it  ``as is''   on  his replica. He will rather  execute a
variant  of $o$,  denoted by  $o'$ (called  a \emph{transformation} of
$o$) that  \textit{intuitively intends to achieve the  same effect as
  $o$}.   This approach  is based  on  a transformation function  $IT$
that applies to couples of concurrent operations defined on the same state.

\begin{example}
  In Figure~\ref{fig:transf},  we illustrate the effect of  $IT$ on the
  previous example.  When $o_2$ is received on site $1$, $o_2$ needs
  to    be    transformed   according    to    $o_1$   as    follows:
  $IT((Del(5),Ins(1,\mbox{\emph{f}}))  =  Del(6)$.  The  deletion
  position  of $o_2$  is incremented  because $o_1$  has  inserted a
  character at position $1$, which  is before the character deleted by
  $o_2$. Next,  $o'_2$ is  executed on site  $1$.  In the  same way,
  when $o_1$ is  received on site $2$, it  is transformed as follows:
  $IT(Ins(1,\mbox{\emph{f}}),Del(5))                             =
  Ins(1,\mbox{\emph{f}})$;  $o_1$   remains  the  same  because
  \emph{f} is inserted before the deletion position of $o_2$.
\end{example}

\subsection{Transformation Properties}

\begin{definition}\label{Def:Extt}
Let $seq$  be  a sequence  of operations.
Transforming  any editing operation  $o$ according to $seq$ is  denoted by
$IT^*(o,seq)$ and is recursively defined as follows:
\begin{gather*}
IT^*(o,[])=o\mbox{ where } [] \mbox{ is the empty sequence;}\\
IT^*(o,[o_1;o_2;\ldots;o_n])=IT^*(IT(o,o_1),[o_2;\ldots;o_n])
\end{gather*}
We say that $o$ has been concurrently generated according to all
operations of $seq$.
\end{definition}

Using   an   OT   algorithm   requires   us to   satisfy   two
properties~\cite{Ressel.ea:96}.
For all $o$, $o_1$ and $o_2$ pairwise concurrent operations:
\begin{itemize}[$\bullet$]
\item \textbf{Condition $TP1$}: $[o_1\,;IT(o_2,o_1)]\,\equiv\, [o_2\,; IT(o_1,o_2)]$.
\item \textbf{Condition $TP2$}:
       $IT^*(o, [o_1\,;IT(o_2,o_1)])\,=\,IT^*(o, [o_2\,; IT(o_1,o_2)])$.
\end{itemize}

Property $TP1$  defines a  \emph{state identity} and ensures that  if $o_1$
and $o_2$ are concurrent, the effect of executing $o_1$ before $o_2$ is
the  same  as  executing  $o_2$  before  $o_1$.  This  property is
necessary but not sufficient  when the number of concurrent operations
is greater than two.
As for $TP2$, it  ensures that transforming $o$
along equivalent and different  operation sequences will give the same
operation.

Properties $TP1$  and $TP2$ are  sufficient to ensure  the convergence
for \textit{any number} of concurrent operations which can be
executed in \textit{arbitrary order}~\cite{Ressel.ea:96}. Accordingly, by these
properties, it is not necessary to enforce a global total order
between concurrent operations because data divergence can always be
repaired by operational transformation. However, finding an IT algorithm that satisfies
$TP1$ and $TP2$ is considered as a hard task, because this proof is often
unmanageably complicated.

%% Verifying that  a given  OT algorithm
%% satisfies $TP1$  and $TP2$ is a computationally  expensive problem even
%% for a  simple text document. Using a  theorem prover to  automate the
%% verification  process  is needed  and  would  be  a crucial  step  for
%% building    correct     collaborative    objects    based     on    OT
%% approach~\cite{imine02b,Imi04}.

%% Using   a  theorem-proving   approach~\cite{imine02b,Imi04},   we  have
%% detected that the function $IT$ of Figure~\ref{fig:grove} contains some
%% subtle bugs that  lead to divergence situations. These
%% situations are detailed in the following section.

\subsection{Partial concurrency problem}

\begin{definition}\label{Def:PaCo}
Two concurrent operations $o_1$ and $o_2$ are said to be
\emph{partially concurrent} iff $o_1$ is generated on list state $l_1$
at site $1$ and $o_2$ is generated on list state $l_2$ at site $2$
with $l_1\neq l_2$.
\end{definition}

In case of partial concurrency situation the  transformation
function $IT$ may lead to data divergence. The following
example illustrates this situation.

\begin{figure}[t]%[htbp]
 \begin{minipage}[t]{0.5\linewidth}
\begin{scriptsize}
\centerline{\xymatrix@C=10pt@M=2pt@R=10pt{
*+[F-,]\txt{site 1 \\ ``fect''} \ar@{.}'[d]'[dd]'[ddd][dddddd] &
*+[F-,]\txt{site 2 \\ ``fect''} \ar@{.}'[d]'[dd]'[ddd][dddddd] \\
o_1=Ins(0,a) \ar[ddr]  & o_3=Ins(0,e) \ar[ddddl] \\
*+[F]{\txt{``afect''}} & *+[F]{\txt{``efect''}} \\
 o_2=Ins(1,f) \ar[ddr]    & IT(o_1,o_3)=Ins(0,a) \\
*+[F]{\txt{``affect''}}  & *+[F]{\txt{``aefect''}} \\
o'_3=Ins(2,e)     & \txt{$o'_2=IT(o_2,o_3)$\\$=Ins(2,f)$} \\
*+[F]{\txt{``afefect''}}  & *+[F]{\txt{``aeffect''}} \\
}}
\end{scriptsize}
  \caption{Wrong application of $IT$.}
  \label{fig:partial_incons}
 \end{minipage}
 \begin{minipage}[t]{0.5\linewidth}
\begin{scriptsize}
\centerline{\xymatrix@C=10pt@M=2pt@R=10pt{
*+[F-,]\txt{site 1 \\ ``fect''} \ar@{.}'[d]'[dd]'[ddd][dddddd] &
*+[F-,]\txt{site 2 \\ ``fect''} \ar@{.}'[d]'[dd]'[ddd][dddddd] \\
o_1=Ins(0,a) \ar[ddr]  & o_3=Ins(0,e) \ar[ddddl] \\
*+[F]{\txt{``afect''}} & *+[F]{\txt{``efect''}} \\
 o_2=Ins(1,f) \ar[ddr]    & IT(o_1,o_3)=Ins(0,a) \\
*+[F]{\txt{``affect''}}  & *+[F]{\txt{``aefect''}} \\
o'_3=Ins(2,e)     & \txt{$o'_2=IT(o_2,IT(o_3,o_1))$\\$=Ins(1,f)$} \\
*+[F]{\txt{``afefect''}}  & *+[F]{\txt{``afefect''}} \\
}}
\end{scriptsize}
  \caption{Correct application of $IT$.}
  \label{fig:partial_correct}
 \end{minipage}

\end{figure}

\begin{example}
Consider two users trying to correct the word ``\emph{fect}'' as in
Figure~\ref{fig:partial_incons}. User $1$ generates two operations
$o_1$ and $o_2$. User $2$ concurrently generates operation
$o_3$. We have $o_1\rightarrow o_2$ and $o_1\parallel o_3$,
but $o_2$ and $o_3$ are partially concurrent as they are generated
on different text states.
At site $1$, $o_3$ has to be transformed against the sequence
$[o_1 ; o_2]$,
\textit{i.e.} $o'_3 = T^*(o_3,[o_1 ; o_2])=Ins(2,e)$. The
execution of $o'_3$ gives the word ``\emph{afefect}''.
At site $2$, transforming $o_1$ against $o_3$ gives
$o'_1 = o_1 = Ins(0,a)$ and transforming $o_2$ against $o_3$ results
in $o'_2=Ins(2,f)$ whose execution leads to the word
``\emph{aeffect}''
which is different from what is obtained at site $1$.
This divergence situation is due to a wrong application of $IT$
to the operations $o_2$ and $o_3$ at site $2$. Indeed, the function
$IT$ requires that both operations are concurrent and defined on
the same state. However, $o_3$ is generated on ``\emph{fect}'' while $o_2$
is generated on ``\emph{afect}''.
\end{example}

In order to solve this partial concurrency problem, $o_2$ should
not be directly transformed with respect to $o_3$ because $o_2$
causally depends on $o_1$ (see Figure~\ref{fig:partial_correct}).
Instead $o_3$ must be transformed against $o_1$ and next $o_2$ may
be transformed against the result.

\subsection{Consistency criteria}

A stable state in an OT-based distributed collaborative editing system is achieved when all
generated operations have been performed at all sites. Thus, the
following criteria must be ensured~\cite{Ellis89,Ressel.ea:96,Sun98}:

\begin{definition}\textbf{(Consistency Model)}\label{def:crit}
An OT-based collaborative editing system is \emph{consistent} iff it satisfies the
following properties:%\vspace{-3mm}
\begin{enumerate}
\item \emph{Causality preservation:}
%for every pair of updates $u_1$ and $u_2$,
if $o_1 \rightarrow o_2$ then $o_1$ is
  executed before $o_2$ at all sites.%\vspace{-3mm}
\item \emph{Convergence:} when all sites have performed the same set
  of updates, the copies of the shared document are identical.
\end{enumerate}
\end{definition}

To preserve the causal dependency between updates, timestamp vectors
are used. The concurrent operations are serialized by using IT
algorithm. As this technique enables concurrent operations to be
serialized in any order, the convergence depends on $TP1$ and $TP2$
that IT algorithm must hold.

\subsection{Integration algorithms}
\label{sec:integration} In   the  OT   approach,  every   site is
equipped by two main components~\cite{Ellis89,Ressel.ea:96}: the
\textit{integration component} and the \textit{transformation
component}.  The integration component is responsible for
receiving, broadcasting and executing operations.  It is  rather
\textit{independent} of the type of the  shared objects. Several
integration algorithms  have been proposed   in   the    groupware
research area,   such as dOPT~\cite{Ellis89},
adOPTed~\cite{Ressel.ea:96},      SOCT2,4
\cite{Suleiman.ea:98,Vidot.ea:00}     and GOTO~\cite{Sun98}.   The
transformation component  is  commonly a  set of  IT algorithms
which is responsible  for merging two concurrent operations
defined on the same state.  Every IT algorithm is
\textit{specific} to the semantics of  a shared object.
%(text in  our example).
Every site generates operations sequentially and stores these operations in a
stack also called a \textit{history} (or \emph{execution trace}).
When a site receives a remote
operation  $o$,  the  integration  component executes  the  following
steps:
\begin{enumerate}
\item From the local history $seq$ it determines the equivalent
      sequence $seq'$ that is the concatenation of two sequences
      $seq_h$ and $seq_c$ where (i) $seq_h$ contains all operations
      happened before $o$ (according to Definition~\ref{Def:caus}), and; (ii)
      $seq_c$ consists of operations that are concurrent to $o$.
    \item It calls the transformation component in order to get
    operation $o'$ that is the transformation of $o$ according to
    $seq_c$ (\textit{i.e.} $o'=IT^*(o,seq_c)$).
    \item It executes $o'$ on the current state.
    \item It adds $o'$ to local history $seq$.
\end{enumerate}

The integration algorithm allows history of executed operations to be
built on every site, provided that the causality relation is
preserved. At stable state, history sites are not necessarily
identical because the concurrent operations may be executed in
different orders. Nevertheless, these histories must be equivalent in
the sense that they must lead to the same final state. This equivalence is
ensured iff the used IT algorithms satisfy properties $TP1$ and $TP2$.

\subsection{State-of-the art transformation algorithms}
In this section, we will present the main IT algorithms known in the
literature for synchronizing linear objects.

\subsubsection{Ellis's algorithm}
Ellis and Gibbs~\cite{Ellis89} are the pioneers of OT approach.
They proposed an IT algorithm to synchronize a shared  text object, shared
by two or more users.  There are two editing operations: $Ins(p,c,pr)$
to insert a character $c$ at  position $p$ and $Del(p,pr)$ to delete a
character at  position $p$.  Operations  $Ins$ and $Del$  are extended
with  another  parameter   $pr$\footnote{This  priority  is
  calculated at the originating site. Two  operations
  generated from  different sites have always  different priorities.}.
This one represents a priority scheme that is used to solve a
conflict occurring  when  two  concurrent  insert  operations were
originally intended  to insert  different characters  at the same
position. Note that concurrent editing operations have always
different priorities.

Algorithm~\ref{fig:grove} gives the four transformation
cases  for $Ins$  and  $Del$  proposed by Ellis and Gibbs. There
are  two interesting situations in the first case ($Ins$ and
$Ins$). The first situation is when the arguments of the two
insert operations are equal (\textit{i.e.} $p_1 = p_2$ and $c_1 =
c_2$).  In this case  the function $IT$ returns the idle operation
$Nop$  that   has  a null  effect on  a text state~\footnote{The
definition of $IT$ is completed by: $IT(Nop,o)=Nop$
  and   $IT(o,Nop)=o$  for  every   operation  $o$.}.    The  second
interesting situation  is when only the insertion  positions are equal
(\textit{i.e.}  $p_1  = p_2$ but $c_1 \neq c_2$).  Such  conflicts are
resolved  by using
the  priority  order  associated  with  each  insert  operation.   The
insertion position will be shifted to the right and will be ($p_1 + 1$) when $Ins$
has a higher priority.  The remaining cases for $IT$ are quite simple.

\begin{algorithm}[h!]
\caption{IT algorithm defined by Ellis and Gibb.}
\label{fig:grove}
\begin{algorithmic}%[1]
\STATE {$IT(Ins(p_1,c_1,pr_1)$, $Ins(p_2,c_2,pr_2))$ =}
\IF{($p_1<p_2$)} {\RETURN $Ins(p_1,c_1,pr_1)$}
\ELSIF {($p_1 > p_2$)} {\RETURN $Ins(p_1+1,c_1,pr_1)$}
       \ELSIF {($c_1 == c_2$)} \RETURN $Nop()$
              \ELSIF {$pr_1 > pr_2$()} \RETURN $Ins(p_1+1,c_1,pr_1)$
              \ELSE \RETURN $Ins(p_1,c_1,pr_1)$
\ENDIF

\medskip
\STATE {$IT(Ins(p_1,c_1,pr_1), Del(p_2,pr_2))$ =}
\IF {($p_1 < p_2$)} \RETURN $Ins(p_1,c_1,pr_1)$
\ELSE \RETURN $Ins(p_1-1,c_1,pr_1)$
\ENDIF

\medskip
\STATE {$IT(Del(p_1,pr_1),Ins(p_2,c_2,pr_2))$ =}
\IF {($p_1<p_2$)} \RETURN $Del(p_1,pr_1)$
\ELSE \RETURN $Del(p_1+1,pr_1)$
\ENDIF

\medskip
\STATE {$IT(Del(p_1,pr_1),Del(p_2,pr_2))$ =}
\IF {($p_1<p_2$)} \RETURN $Del(p_1,pr_1)$
\ELSIF {($p_1>p_2$)} \RETURN $Del(p_1-1,pr_1)$
\ELSE \RETURN  $Nop()$
\ENDIF
\end{algorithmic}
\end{algorithm}

\subsubsection{Ressel's algorithm}
Ressel et \textit{al.}~\cite{Ressel.ea:96} proposed an algorithm that
provides two modifications in Ellis's algorithm in order to satisfy
properties $TP1$ and $TP2$. The first modification consists in
replacing priority parameter $pr$ by another parameter $u$, which is
simply the \emph{identifier} of the issuer site. Similarly, $u$ is
used for tie-breaking when a conflict occurs between two concurrent
insert operations.

As for the second modification, it concerns how a pair of  insert
operations is transformed. When two concurrent insert operations add
at the same position two (identical or different) elements, only the
insertion position of operation having a higher identifier is
incremented. In other words, the both elements are inserted even if
they are identical. What is opposite to solution proposed by Ellis and
Gibbs, which keeps only one element in case of identical concurrent
insertions. Apart from these modifications, the other cases remain
similar to those of Ellis and Gibb.
Algorithm~\ref{alg:resselalgo} gives all transformation cases proposed
by Ressel et \textit{al.}~\cite{Ressel.ea:96}.

\begin{algorithm}[h!]
\caption{IT algorithm defined by Ressel and \textit{al}.}
\label{alg:resselalgo}
\begin{algorithmic}%[1]
\STATE {$IT(Ins(p_1,c_1,u_1),Ins(p_2,c_2,u_2))$ =}
\IF {($p_1<p_2$ or ($p_1=p_2$ and $u_1<u_2$))} \RETURN $Ins(p_1,c_1,u_1)$
\ELSE \RETURN $Ins(p_1+1,c_1,u_1)$
\ENDIF

\medskip
\STATE {$IT(Ins(p_1,c_1,u_1), Del(p_2,u_2))$ =}
\IF {($p_1 \leq p_2$)}
\RETURN $Ins(p_1,c_1,u_1)$
\ELSE \RETURN $Ins(p_1-1,c_1,u_1)$
\ENDIF

\medskip
\STATE {$IT(Del(p_1,u_1),Ins(p_2,c_2,u_2))$ =}
\IF {($p_1<p_2$)}
\RETURN $Del(p_1,u_1)$
\ELSE \RETURN $Del(p_1+1,u_1)$
\ENDIF

\medskip
\STATE {$IT(Del(p_1,u_1),Del(p_2,u_2))$ =}
\IF {($p_1<p_2$)}
\RETURN $Del(p_1,u_1)$
\ELSIF {($p_1>p_2$)}
\RETURN $Del(p_1-1,u_1)$
\ELSE \RETURN  $Nop()$
\ENDIF
\end{algorithmic}
\end{algorithm}

\subsubsection{Sun's algorithm}
Sun et \textit{al}.~\cite{Sun.ea:98} proposed another solution
as given in Algorithm~\ref{alg:sunalgo}. This algorithm is
slightly different in the sense that it is defined for stringwise
operations. Indeed, the following operations are used:
\begin{itemize}
\item $Ins(p,s,l)$: insert string $s$ of length $l$ at position $p$;
\item $Del(p,l)$: delete string of length $l$ from position $p$.
\end{itemize}

For instance, to apply the inclusion transformation to operation $o_1=Ins(p_1,s_1,l_1)$
against operation $o_2=Del(p_2,l_2)$ , if $p_1\leq p_2$,
then $o_1$ must refer to a position which is to the left of or at the
position referred to by $o_2$,
so the assumed execution of $o_2$ should not have any impact on the
intended position of $o_1$.
Therefore, no adjustment needs to be made to $o_1$. However, if $p_1 > (p_2+l_2)$,
which means that the position of $o_1$ goes beyond the rightmost position in
the deleting range of $o_2$, the intended position of $o_1$ would
have been shifted by $l_2$ characters to the left if the impact of
executing $o_2$ was taken into account. Therefore, the position
parameter of $o_1$ is decremented by $l_2$. Otherwise,
it must be that the intended position of $o_1$ falls in the deleting
range of $o_2$. In this case, $o_2$ should not delete any characters
to be inserted by $o_1$, and the new inserting position should be $p_2$.

\begin{algorithm}[h!]
\caption{Stringwise IT algorithm defined by Sun and \textit{al}.}
\label{alg:sunalgo}
\begin{algorithmic}%[1]
\STATE {$IT(Ins(p_1,s_1,l_1)$, $Ins(p_2,s_2,l_2))$ =}
\IF {($p_1<p_2$)} \RETURN $Ins(p_1,s_1,l_1)$
\ELSE \RETURN $Ins(p_1+1,s_1,l_1)$
\ENDIF

\medskip
\STATE {$IT(Ins(p_1,s_1,l_1)$, $Del(p_2,l_2))$ =}
\IF {($p_1 \leq p_2$)} \RETURN $Ins(p_1,s_1,l_1)$
\ELSIF{($p_1 > p_2 + l_2$)} \RETURN $Ins(p_1-l_2,s_1,l_1)$
\ELSE \RETURN $Ins(p_2,s_1,l_1)$
\ENDIF

\medskip
\STATE {$IT(Del(p_1,l_1)$,$Ins(p_2,s_2,l_2))$ =}
\IF {($p_1 \leq p_2$)} \RETURN $Del(p_1,l_1)$
\ELSIF {($p_1 \geq p_2$)} \RETURN $Del(p_1+l_2,l_1)$
\ELSE \RETURN $[Del(p_1,p_2 - p_1) ; Del(p_2 + l_2,l_1 - (p_2 - p_1))]$
\ENDIF

\medskip
\STATE {$IT(Del(p_1,l_1)$,$Del(p_2,l_2))$ =}
\IF {($p_2 \geq p_1 + l_1$)} \RETURN $Del(p_1,l_1)$
\ELSIF {($p_1 \geq p_2 + l_2$)} \RETURN $Del(p_1 - l_2,l_1)$
\ELSIF {($p_2 \leq p_1$ and $p_1+l_1 \leq p_2 + l_2$)} \RETURN $Del(p_1,0)$
\ELSIF {($p_2 \leq p_1$ and $p_1+l_1 > p_2 + l_2$)} \RETURN $Del(p_2,(p_1+l_1)-(p_2+l_2))$
\ELSIF {($p_2 > p_1$ and $p_2+l_2 \geq p_1 + l_1$)} \RETURN $Del(p_1,p_2 - p_1)$
\ELSE \RETURN  $Del(p_1,l_1 - l_2)$
\ENDIF
\end{algorithmic}
\end{algorithm}

To better compare with other IT algorithms, we have transformed the
proposition of Sun and \textit{al}. into elementwise (or
characterwise) one (see Algorithm~\ref{alg:sun1algo}).

\begin{algorithm}[h!]
\caption{Characterwise IT algorithm of Sun and \textit{al}.}
\label{alg:sun1algo}
\begin{algorithmic}%[1]
\STATE {$IT(Ins(p_1,c_1)$, $Ins(p_2,c_2))$ =}
\IF {($p_1<p_2$)} \RETURN $Ins(p_1,c_1)$
\ELSE \RETURN $Ins(p_1 + 1,c_1)$
\ENDIF

\medskip
\STATE {$IT(Ins(p_1,c_1)$, $Del(p_2))$ =}
\IF {($p_1 \leq p_2$)} \RETURN $Ins(p_1,c_1)$
\ELSE \RETURN $Ins(p_1-1,c_1)$
\ENDIF

\medskip
\STATE {$IT(Del(p_1)$,$Ins(p_2,c_2))$ =}
\IF {($p_1<p_2$)} \RETURN $Del(p_1)$
\ELSE \RETURN $Del(p_1+1)$
\ENDIF

\medskip
\STATE {$IT(Del(p_1)$,$Del(p_2))$ =}
\IF {($p_1<p_2$)} \RETURN $Del(p_1)$
\ELSIF {($p_1>p_2$)} \RETURN $Del(p_1-1)$
\ELSE \RETURN  $Nop()$
\ENDIF
\end{algorithmic}
\end{algorithm}

\subsubsection{Suleiman's algorithm}
Suleiman and \textit{al}.~\cite{suleiman97} proposed another
transformation algorithm
that modifies the signature of insert operation by adding two
parameters $av$ and $ap$. These parameters store the set of concurrent
delete operations. For an insert operation $Ins(p,c,av,ap)$,
$av$ contains operations that
have deleted a character before the insertion position $p$. As for
$ap$, it contains operations that have removed a character after $p$.
When an insert operation is generated the parameters $av$ and $ap$ are
empty. They will be filled during transformation steps.

All transformation cases of  Suleiman et \textit{al}. are given in
Algorithm~\ref{alg:suleimanalgo}. To resolve the conflict between two
concurrent insert operations $Ins(p,c_1,av_1,ap_1)$ and
$Ins(p,c_2,av_2,ap_2)$, three cases are possible:
\begin{enumerate}
\item $(av_1 \cap ap_2) \neq \emptyset$: character $c_2$ is inserted before
  character $c_1$,
\item $(ap_1 \cap av_2) \neq \emptyset$: character $c_2$ is inserted after
character $c_1$,
\item $(av_1 \cap ap_2) = (ap_1 \cap av_2) = \emptyset$: in this case
  function $code(c)$, which computes a total order on characters
  (\textit{e.g.} lexicographic order), is used to choose among $c_1$
  and $c_2$ the character to be added before the other. Like the site
  identifiers and priorities, $code(c)$ enables us to tie-break
  conflict situations.
\end{enumerate}

Note that when two concurrent operations insert the same character
(\textit{e.g.} $code(c_1)=code(c_2)$) at the same position, the one is
executed and the other one is ignored by returning the idle operation $Nop$.
In other words, like the solution of Ellis and Gibb~\cite{Ellis89},
only one character is kept.

\begin{algorithm}[h!]
\caption{IT algorithm of Suleiman and \textit{al}.}
\label{alg:suleimanalgo}
\begin{algorithmic}%[1]
\STATE {$IT(Ins(p_1,c_1,av_1,ap_1),Ins(p_2,c_2,av_2,ap_2))$ =}
\IF {($p_1<p_2$)} \RETURN $Ins(p_1,c_1,av_1,ap_1)$
\ELSIF {($p_1>p_2$)} \RETURN $Ins(p_1+1,c_1,av_1,ap_1)$
\ELSIF {($av_1\cap ap_2  \neq \emptyset$)} \RETURN  $Ins(p_1+1,c_1,av_1,ap_1)$
\ELSIF {($ap_1\cap av_2  \neq \emptyset$)} \RETURN $Ins(p_1,c_1,av_1,ap_1)$
\ELSIF {($code(c_1)  > code(c_2)$)} \RETURN $Ins(p_1,c_1,av_1,ap_1)$
\ELSIF {($code(c_1) < code(c_2)$)} \RETURN $Ins(p_1+1,c_1,av_1,ap_1)$
\ELSE \RETURN $Nop()$
\ENDIF

\medskip
\STATE {$IT(Ins(p_1,c_1,av_1,ap_1), Del(p_2))$ =}
\IF {($p_1 \leq p_2$)}
\RETURN $Ins(p_1,c_1,av_1,ap_1\cup \{Del(p_2)\})$
\ELSE \RETURN $Ins(p_1-1,c_1,av_1\cup \{Del(p_2)\},ap_1)$
\ENDIF

\medskip
\STATE {$IT(Del(p_1),Ins(p_2,c_2,av_2,ap_2))$ =}
\IF {($p_1<p_2$)}
\RETURN $Del(p_1)$
\ELSE \RETURN $Del(p_1+1)$
\ENDIF

\medskip
\STATE {$IT(Del(p_1),Del(p_2))$ =}
\IF {($p_1<p_2$)}
\RETURN $Del(p_1)$
\ELSIF {($p_1>p_2$)}
\RETURN $Del(p_1-1)$
\ELSE \RETURN  $Nop()$
\ENDIF
\end{algorithmic}
\end{algorithm}

\subsubsection{Imine's algorithm}
In~\cite{imine02b}, Imine and \textit{al}. proposed another IT
algorithms which again enriches the signature of insert
operation. Indeed, they redefined as $Ins(p,ip,c)$ where $p$ is the
current insertion position and $ip$ is the initial (or the original)
insertion position given at the generation stage.
Thus, when transforming a pair of insert operations having the same
current position, they compared first their initial positions in order to
recover the position relation at the generation phase. If the initial
positions are identical, then like Suleiman and
\textit{al}.~\cite{suleiman97} they used function $code(c)$ to tie-break
an eventual conflict.

 The parameters $p$ and $ip$ are initially identical when
the operation is generated. For example, if a user inserts a
character $z$ at position $3$, operation $Ins(3,3, x)$ is
generated. When this operation is transformed, only the current
position (first parameter) will change. The initial position
parameter remains unchanged. Algorithm~\ref{alg:iminealgo} gives
transformation cases.\\

\begin{algorithm}[h!]
\caption{IT algorithm of Imine and \textit{al}.}
\label{alg:iminealgo}
\begin{algorithmic}%[1]
\STATE {$IT(Ins(p_1,o_1,c_1),Ins(p_2,o_2,c_2))$ =}
\IF {($p_1<p_2$)} \RETURN $Ins(p_1,o_1,c_1)$
\ELSIF {($p_1>p_2$)} \RETURN $Ins(p_1+1,o_1,c_1)$
\ELSIF {($o_1<o_2$)} \RETURN  $Ins(p_1,o_1,c_1)$
\ELSIF {($o_1>o_2$)} \RETURN  $Ins(p_1+1,o_1,c_1)$
\ELSIF {($code(c_1) < code(c_2)$)} \RETURN $Ins(p_1,o_1,c_1)$
\ELSIF {($code(c_1) > code(c_2)$)} \RETURN $Ins(p_1+1,o_1,c_1)$
\ELSE \RETURN $Nop()$
\ENDIF

\medskip
\STATE {$IT(Ins(p_1,o_1,c_1), Del(p_2))$ =}
\IF {($p_1 \leq p_2$)}
\RETURN $Ins(p_1,o_1,c_1)$
\ELSE \RETURN $Ins(p_1-1,o_1,c_1)$
\ENDIF

\medskip
\STATE {$IT(Del(p_1),Ins(p_2,o_2,c_2))$ =}
\IF {($p_1<p_2$)}
\RETURN $Del(p_1)$
\ELSE \RETURN $Del(p_1+1)$
\ENDIF

\medskip
\STATE {$IT(Del(p_1),Del(p_2))$ =}
\IF {($p_1<p_2$)}
\RETURN $Del(p_1)$
\ELSIF {($p_1>p_2$)}
\RETURN $Del(p_1-1)$
\ELSE \RETURN  $Nop()$
\ENDIF
\end{algorithmic}
\end{algorithm}

\section{Concrete model}
The following sections are devoted to the specification and
analysis of IT algorithms, by means of model-checker
\emph{UPPAAL}. We show how to exploit some features of IT
algorithms and the specification language of \emph{UPPAAL} to
attenuate the state explosion problem of the execution environment of such
algorithms.

In \emph{UPPAAL}, a system consists of a collection of processes
which can communicate via some shared data and synchronize through
binary or broadcast channels.  Each process is an automaton
extended with finite sets of clocks, variables (bounded integers),
guards and actions. In such automata, locations can be labelled by
clock conditions and edges are annotated with selections, guards,
synchronization signals and updates. Selections
 bind non-deterministically a given identifier to a value in a
given range (type). The other three labels of an edge are within
the scope of this binding. An edge is enabled in a state if and
only if the guard evaluates to true. The update expression of the
edge is evaluated when the edge is fired. The side effect of this
expression changes the state of the system. Edges labelled with
complementary synchronization signals over a common channel must
synchronize. Two or a more processes synchronize through channels
with a sender/receiver syntax \cite{Berard}. For a binary channel,
a sender can emit a signal through a given binary channel $Syn$
($Syn!$), if there is another process (a receiver) ready to
receive the signal ($Syn?$). Both sender and receiver synchronize
on execution of complementary actions $Syn!$ and $Syn?$. For a
broadcast channel, a sender can emit a signal through a given
broadcast channel $Syn$ ( $Syn!$), even if there is no process
ready to receive the signal ($Syn?$). When a sender emits such a
signal via a broadcast channel, it is synchronized with all
processes ready to receive the signal. The updates of synchronized
edges are executed starting with the one of the sender followed by
those of the receiver(s). The execution order of updates of
receivers complies with their creation orders (i.e., if a receiver
$A$ is created before receiver $B$ then the update of $A$ is
executed before the one of $B$).

A replication-based distributed collaborative editing system is composed of
two or more sites (users) which communicate via a network and use
the principle of multiple copies, to share some object (a text).
Initially, each user has a copy of the shared object. It can
afterwards modify its copy by executing operations generated
locally and those received from other users. When a site executes
a local operation, it is broadcast to all other users.

This system is modelled as a set of variables, functions, a
broadcast channel and processes (one per user). Note that the
network is abstracted and not explicitly represented. This is
possible by putting visible (in global variables) all operations
generated by different sites and timestamp vectors of sites. As we
will explain later, in this way, there is no need to represent and
manage queues of messages.

\subsection{Input data and variables}

The system model has the following inputs:\begin{enumerate}
    \item The number of sites $(const \ int \ NbSites)$; Each site has its own identifier,
    denoted $pid$ for process identifier ($pid \in [0, NbSites-1]$).
    \item The initial text to be shared by users and its alphabet. The text to be shared by users
    is supposed to be infinite but the attribute \emph{Position} of operations is restricted
    to the window $[0,L-1]$ of the text. The length
    of the window is set in the constant $L$ ($const \ int \ L$).
    \item The number of local operations of each site, given in array
    $Iter[NbSites]$\\
    ($const \ int \ Iter[NbSites]$, $Iter[i]$ being the number of local operations of site
    $i$).\\ We also use and set in constant named $MaxIter$ the total number of operations ($const \ int \ MaxIter=\underset{i \in
[0,NbSites-1]}\sum Iter[i]$);
    \item The IT algorithm ($const \ int \ algo$).
\end{enumerate}

Variables are of two kinds: those used to store input data and
those used to manage the execution of operations:
\begin{enumerate}
    \item The different copies (one per site) of the shared text are stored in the array
    ($text[NbSites][L]$). Each site $i$ executes operations on its copy of text (i.e.
$text[i]$). To make visible the effect of operations executed
    on different texts, all entries of the text are initialized with
    $-1$. The alphabet considered here may be any bounded interval of non-negative integer
    numbers. This restriction is in fact imposed by the language of
    \emph{UPPAAL} as it does not allow to define strings.
    \item Timestamp vectors of different sites are kept in the array $V[NbSites][NbSites]$.
    \item Vector $Operations[MaxIter]$ is used to store
all operations selected by the different sites and their timestamp
vectors. Each operation has its own identifier, corresponding to
its entry in array $Operations$. Recall that there are two kinds
of operations: \emph{Del} and \emph{Ins}. The Delete operation has
as attribute the \emph{position} in the text of the character to
be deleted. The Insert operation has two attributes
\emph{position} and \emph{character} which indicate the position
where the character has to be inserted. The attribute position of
each operation may be any value inside $[0,L-1]$, $L$ being the
length of the text window to be observed. The attribute
\emph{character} of the insert operation may be any element of the
text alphabet.
    \item Array $List[NbSites][MaxIter]$ is used to save traces and signatures of operations as they are
exactly executed by each site. Recall that before executing a non
local operation, a site may apply some IT algorithm on the
operation. The resulting operation is then executed on its copy of
text. We consider here the five IT algorithms presented in the
previous section: \emph{Ellis}' algorithm, \emph{Ressel}'s
algorithm, \emph{Sun}'s algorithm, \emph{Imine}'s algorithm and
\emph{Suleiman}'s algorithm. The structure of elements of array
$List$ depends, in fact, on the IT algorithm. For algorithms of
\emph{Ellis, Ressel, Sun} and \emph{Imine}, this structure is
composed of the identifier of the operation and the current
position. For the algorithm of \emph{Suleiman}, we need, in
addition, for each insert operation, two vectors to memorize
identifiers of $delete$ operations executed respectively before
and after the operation. Note that structures $trace\_t$ and
$operation\_t$ may need to be redefined for other IT algorithms.
    \item The broadcast channel $Syn$ used for synchronization of some operations
    and also for synchronization on termination.
 \end{enumerate}

Table \ref{decl} gives the above declarations in \emph{UPPAAL}
language. For example, the declaration \emph{int[-1,1]
text[NbSites][L]} defines an array of bounded
integers. Each element of this array is an integer between $-1$ and $1$.
Note that these variables are defined as global to be accessible
by any site (avoiding duplication of data in the representation of
the system state). In addition, this eases the specification of
the convergence property and allows to force the execution, in one
step, some edges of different sites. For example, sites can be
synchronized on termination: when a site completes the execution
of all operations, it stays in the current state until all other
sites complete the execution of all operations. Then, they leave
to reach together their respective termination
states. Therefore, this synchronization allows to reduce the number of reachable states.
Indeed, intermediate states, where some sites are in their termination states and some others
are not, are not accessible with this synchronization. They are however accessible without this
synchronization.\\

\begin{table}[h!]
\begin{center}
\caption{Declaration of constants, types and variables of the
concrete model} \label{decl}
\begin{tabular} {|l|}
\hline
// Declaration of constants \\
\emph{const  int  NbSites = 3};\\
\emph{const  int Iter[NbSites]= \{1,1,1\}}; \\
\emph{const int MaxIter=Iter[0]+Iter[1]+Iter[2]};\\
\emph{const int  L= 2*MaxIter};\\
\emph{const int  Del= 0};\\
\emph{const int  Ins= 0};\\
\emph{const int  Ellis= 0};\\
\emph{const int  Ressel= 1};\\
\emph{const int  Sun= 2};\\
\emph{const int  Suleiman= 3};\\
\emph{const int  Imine= 4};\\
\emph{const int[Ellis,Imine]  algo = Ellis};\\

// Declaration of types\\
\noindent \emph{typedef int[0, NbSites-1] pid\_t;}\\
\emph{typedef int[0, 1] alphabet;}\\
\emph{typedef int[Del, Ins] operator;}\\
\emph{typedef struct \{pid\_t Owner; operator opr; int pos;
alphabet x; int V[NbSites];\} operation\_t;}\\
%\emph{typedef struct \{int numOp; int posC; \} trace\_t ;}\\
\emph{typedef struct \{int numOp; int posC; int a[MaxIter-1]; int b[MaxIter-1]; \} trace\_t ;}\\

// Declaration of variables\\
\noindent \emph{int[-1,1] text[NbSites][L]};\\
\emph{operation\_t Operations[MaxIter];}\\
\emph{int[0,MaxIter-1] V[NbSites][MaxIter];}\\
\emph{trace\_t List[NbSites][MaxIter];}\\
\emph{int[0,MaxIter] ns;}\\

// Declaration of a broadcast channel\\
\noindent \emph{broadcast channel Syn;}\\
\hline
\end{tabular}
\end{center}
\end{table}

\subsection{Behavior of each site}

\begin{figure}[h!]
\begin{center}
\noindent
\includegraphics[width=0.70\textwidth]{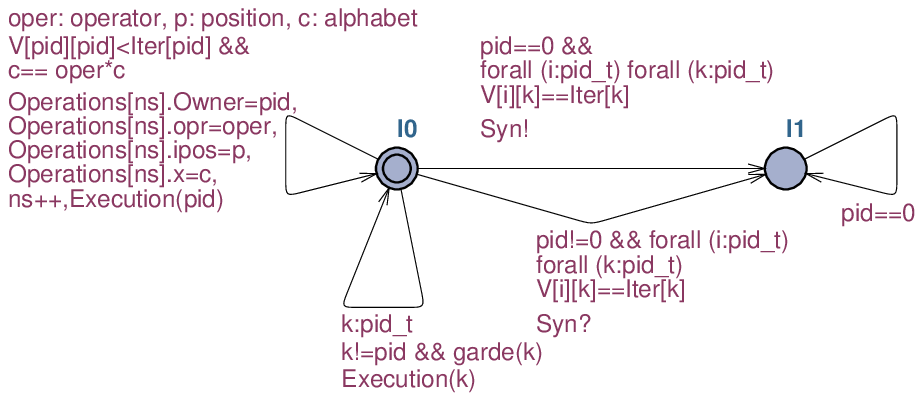}
\caption{The concrete model} \label{ConcreteModel}
\end{center}
\end{figure}

Behaviors of sites are similar and represented by a type of
process named $Site$. The process behavior of each site is
depicted by the automaton shown in Figure \ref{ConcreteModel}. The
only parameter of the process is the site
identifier named $pid$.

Using \emph{UPPAAL}, the definition of the system is given by the
following declarations which mean that the system consists of
$NbSites$ sites of type $Site$:

\noindent \emph{Sites(const pid\_t pid) = Site(pid);}\\
\noindent \emph{system \ Sites; }\\

Each user executes, one by one, all operations (local and non
local ones), on its own copy of the shared text
(loops on location $l0$ of Figure \ref{ConcreteModel}).
The execution order of operations must, however, respect the
causality principle. The causality principle is ensured by the
timestamp vectors of sites $V[NbSites][NbSites]$. For each pair of
sites $(i,j)$, element $V[i][j]$ is the number of operations of
site $j$ executed by site $i$. $V[i][i]$ is then the number of
local operations executed in site $i$. Note that $V[i][j]$ is also
the rank of the next operation of site $j$ to be executed by site $i$.
Timestamp vectors are also used, in the IT algorithms,
to determine whether operations are concurrent or dependent.
Initially, entries of the timestamp vector of every site $i$ are
set to $0$. Afterwards, when site $i$ executes an operation of a
site $j$ ($j \in [0,NbSites-1]$), it
increments the entry of $j$ in its own timestamp vector (i.e $V[i][j]++$).

\subsubsection{Execution of a local operation}

A local operation can be selected and executed by a site $pid$ if
the number of local operations already executed by site $pid$ does
not yet reach its maximal number of local operations (i.e.
$V[pid][pid] < Iter[pid]$). In this case, its timestamp vector is
set to the timestamp vector of its site. The owner, the signature
and the timestamp vector of the selected operation are stored in
array $Operations$. The execution of the operation consists of
calling function $Operation$ (see functions $Execution$ and
$Operation$ in the Appendix). Its broadcast to other sites is
simply simulated by incrementing the number of local operations
executed by the sender site ($V[pid][pid]++$). The execution of a
local operation is represented by the loop on location $l0$ which
consists of $3$ parts: the selection of an operation ($oper:
operator, p: position, c: alphabet$), the guard
$(V[pid][pid]<Iter[pid] \ \&\& \ c== oper*c)$ and the update
$(Operations[ns].ipos=p, Operations[ns].x=c, ns++,
Execution(pid))$. The second part of the guard imposes that, for
the delete operation, the argument character is always set to $0$.
The update part stores and executes the selected operation.

Initially, every site $pid$ is in its initial location $l0$,
entries of $V[pid]$ are set to $0$ and, for instance,
$Iter[pid]=1$. For this initial state, only the loop corresponding
to the execution of a local operation is multi-enabled (one
enabling for each operation signature which satisfies the guard of
the loop). For example, for $oper=Ins, \ p=2$ and $c=1$, the
execution of this loop leads to another state with the same
location but different values of variables. Indeed, when this edge
is executed the selected operation is stored in $Operations$ and
executed on the local copy of the text (see the code of function
$Execution$ in the Appendix).

\subsubsection{Execution of a non local operation}

A site $pid$ can execute an operation of another site $k$ if there
is an operation of $k$ executed by $k$ but not yet executed by
$pid$ (i.e.: $V[pid][k] < V[k][k]$) and its timestamp vector is
less or equal to the timestamp vector of site $pid$ (i.e.:
$\forall j \in [0,NbSites-1], V[pid][j] >= Operations[num].V[j]$,
$num$ being the identifier of the operation). Before executing a
non local operation, it may be transformed using a given
IT algorithm (see functions $garde$, $Execution$ and
$Operation$ in the Appendix). The execution of a non local
operation is represented by the loop on location $l0$ which
consists of $3$ parts: the selection of a site ($k:pid\_t$), the
guard $(k!=pid \ \&\& \ garde(k))$ and the update $Execution(k)$.
The first and second parts select a non local operation w.r.t the
causality principle. The update part
 executes the integration steps for the selected operation
 according with explanation given in section \ref{sec:integration}.
 Note that the partial concurrency problem (see section \ref{sec:partialconcurrency})
 is also treated in function $Execution$.

\subsubsection{Termination of different sites}
When a site completes the execution of all operations, it waits
for the termination of all other sites. The synchronization on
termination is realized by means of the broadcast channel $Syn$
and all edges connecting locations $l0$ and $l1$. The site $0$ is
the sender of signals $Syn$ and all others are receivers of
signals $Syn$.\\

\subsection{Convergence property} The main required property for
the system is the convergence property. This property states that
whenever two sites complete the execution of the same set of
operations, their resulting texts must be identical. To specify
this property, we define the notion of stable state. A stable
state of the system is a situation where all sent operations are
received and executed (there is no operation in transit). A site
$i$ is in a stable state if all operations sent to site $i$ are
received and executed by $i$ (i.e. $ forall (k:pid\_t) \
V[i][k]==V[k][k]$). The convergence property can be rewritten
using the notion of stable state as follows: "Whenever two sites
$i$ and $j$ are in stable state, they have identical texts". This
can be also specified by
the following \emph{UPPAAL}'s $CTL$ formula:
$$A \Box\ (exists (i:pid\_t) \ exists (j:pid\_t) $$
$$ i!=j \ {\&\&} \ forall (k:pid\_t) \ V[i][k]==V[k][k] \ {\&\&} \  V[j][k]==V[k][k])$$
$$\ imply \ forall (l: int[0,L-1]) \ text[i][l]==text[j][l]$$

%% $$A[] \ (exists (i:pid\_t) \ exists (j:pid\_t) $$
%% $$ i!=j \ {\&\&} \ forall (k:pid\_t) \ V[i][k]==V[k][k] \ {\&\&} \  V[j][k]==V[k][k])$$
%% $$\ imply \ forall (l:pid\_t) \ text[i][l]==text[j][l]$$

This formula means that for each execution path and for each state
of the execution path if any two sites $i$ and $j$ are in stable
states then their copies of text $text[i]$ and $text[j]$ are
identical. We consider, in the following the negation of the above
formula, referred by $\phi_1$:

$$E\Diamond  \ (exists (i:pid\_t) \ exists (j:pid\_t) $$ $$ i!=j \ \&\& \ forall
(k:pid\_t) \ V[i][k]==V[k][k] \ \&\& \ V[j][k]==V[k][k]) $$ $$
\&\& \ exists (l: int[0,L-1])  \ text[i][l]!=text[j][l]$$

%% $$E<>  \ (exists (i:pid\_t) \ exists (j:pid\_t) $$ $$ i!=j \ \&\& \ forall
%% (k:pid\_t) \ V[i][k]==V[k][k] \ \&\& \ V[j][k]==V[k][k]) $$ $$
%% \&\& \ exists (l:pid\_t)  \ text[i][l]!=text[j][l]$$

\subsection{Verification of properties}

We have tested the five transformation algorithms considered here,
using the concrete model. We report in, Table
\ref{TableConcreteModel}, results obtained, in case of 3 sites
($NbSites=3$), 3 or 4 operations ($MaxIter=3$ or $MaxIter=4$), and
a window of the observed text of length $L=2*MaxIter$, for two
properties: the negation of the convergence property ($\phi_1$)
and the absence of deadlocks ($\phi_2:$ $A\Box \ not \ Deadlock$).
A state $q$ of a model is in deadlock if and only if there is no
action enabled in $q$ nor in states reachable from $q$ by time
progression. Note that we use the above input data for all tested
models and all tests are performed using the version 4.0.6 of
UPPAAL 2k on a 3 Gigahertz {Pentium-4} with 1GB of RAM.

We give, in column 4, for each algorithm and each property, the
number of explored, the number of computed states, and the
execution time (CPU time in seconds). Note that for 3 sites, the
verification of $\phi_2$ was aborted for a lack of memory. We have
encountered the same problem for 4 sites and formula $\phi_1$. The
first property is always satisfied and allows us to compute the
size of the entire state space. The second one is satisfied for
algorithms of $Ellis$, $Ressel$ and $Sun$ but not satisfied for
$Imine$'s and $Suleiman$'s algorithms.

\begin{table}[h!]
\begin{center}
\caption{Concrete model} \label{TableConcreteModel} \vspace{2pt}
\begin{tabular}{|c||c||c||c|}
\hline $Alg. \ NbSites \ MaxIter $ & Prop. &Val.& Expl. / Comp. /
Time (s)
\\ \hline
 $Ellis \ 3 \ 3 $ & $\phi_1$  & true & 825112 / 1838500 / 121.35 \\
  $Ellis \ 3 \ 3 $& $\phi_2$ & ? & ?  \\ \hline
$Ressel \ 3 \ 3$ & $\phi_1$  & true & 833558 / 1851350 / 122.76 \\
$Ressel \ 3 \ 3$  & $\phi_2$ & ? & ?  \\ \hline
$Sun \ 3 \ 3$ & $\phi_1$  & true & 836564 / 1897392 / 122.33  \\
$Sun \ 3 \ 3$ & $\phi_2$ & ? & ? \\ \hline
$Suleiman \ 3 \ 3$ & $\phi_1$  &false& 3733688 / 3733688 / 365.06  \\
$Suleiman \ 3 \ 3$  & $\phi_2$ & ? & ?   \\
$Suleiman \ 3 \ 4$ & $\phi_1$  & ? & ? \\
\hline
$Imine \ 3 \ 3$ & $\phi_1$  & false & 3733688 / 3733688 /  361.16\\
$Imine \ 3 \ 3$ & $\phi_2$ & ? &  ? \\
$Imine \ 3 \ 4$ & $\phi_1$  & ? &  ? \\
\hline
\end{tabular}
\end{center}
\end{table}

%% \begin{table}
%% \begin{center}
%% \caption{Case of the concrete model} \label{TableConcreteModel}
%% \vspace{2pt}
%% \begin{tabular}{|c||c||c||c|}
%% \hline $Alg.$ & Prop. &Val.& Expl. / Comp. / Time (s)
%% \\ \hline
%%  $Ellis$ & $\phi_1$  & true & 825112 / 1838500 / 121.35 \\
%%  $3 \ sites$ & $\phi_2$ & ? & ?  \\ \hline
%% $Ressel$ & $\phi_1$  & true & 833558 / 1851350 / 122.76 \\
%%   $3 \ sites$ & $\phi_2$ & ? & ?  \\ \hline
%% $Sun$ & $\phi_1$  & true & 836564 / 1897392 / 122.33  \\
%% $3 \ sites$ & $\phi_2$ & ? & ? \\ \hline
%% $Suleiman$ & $\phi_1$  &false& 3733688 / 3733688 / 365.06  \\
%% $3 \ sites$ & $\phi_2$ & ? & ?   \\
%% $4 \ sites$ & $\phi_1$  & ? & ? \\
%% \hline
%% $Imine$ & $\phi_1$  & false & 3733688 / 3733688 /  361.16\\
%% $3 \ sites$ & $\phi_2$ & ? &  ? \\
%% $4 \ sites$ & $\phi_1$  & ? &  ? \\
%% \hline
%% \end{tabular}
%% \end{center}
%% \end{table}

As an example, we report, in Table \ref{Traces}, for Ellis's
algorithm, the execution traces of sites, given by \emph{UPPAAL},
which violate the convergence property. It corresponds to the case
where the local operations of sites 0, 1 and 2 are respectively
$Del(0)$, $Ins(0,0)$ and $Ins(1,0)$ respectively numbered $0$, $1$
and $2$. The convergence property is violated for sites $0$ and
$1$ when the execution orders of these operations are $0 \ 1 \ 2$
in site $0$ and $1 \ 0 \ 2$ in site $1$.

\begin{table}[h!]
\begin{center}
\caption{Execution traces violating the convergence property in
case of Ellis's algorithm} \label{Traces} \vspace{2pt}
\begin{tabular}{|l||l|l|l|}
\hline Variables & Site 0 & Site 1 & Site 2
\\ \hline \hline
 $Operations$ & $Del \ 0$ & $Ins \ 0 \ 0$&  $Ins \ 1  \ 0$  \\
 \hline \hline
 $List$ & $0 \ \ Del \ 0$  & $1 \ \ Ins \ 0 \ 0$ & $2 \ \ Ins \ 1 \ 0$ \\
     & $1 \ \ Ins \ 0 \ 0$ & $0 \ \ Del \ 1 $ &   \\
     & $2 \ \ Nop$ & $2 \ \ Ins \ 1 \ 0 $ &   \\ \hline
 $text$ & $0 \ -1 \ ...$ & $ 0 \ 0 \ -1 \ ...$ & $ -1 \ 0 \ ...$ \\ \hline
\end{tabular}
\end{center}
\end{table}

\subsection{Preselecting signatures of operations}

\begin{figure}[h!]
\begin{center}
\includegraphics[width=0.60\textwidth]{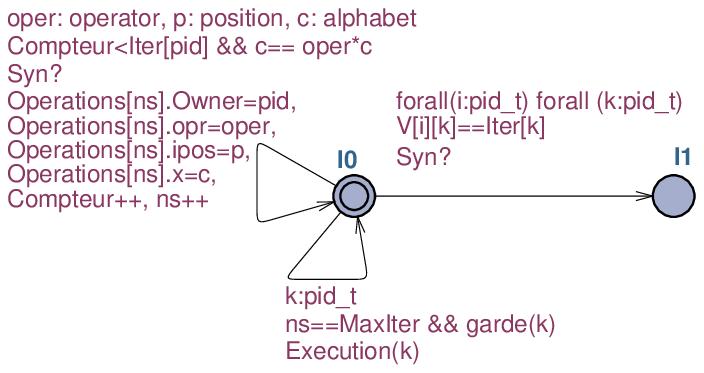}
\caption{Variant 1 of the concrete model: Process Site}
\label{Preselection1}
\includegraphics[width=0.65\textwidth]{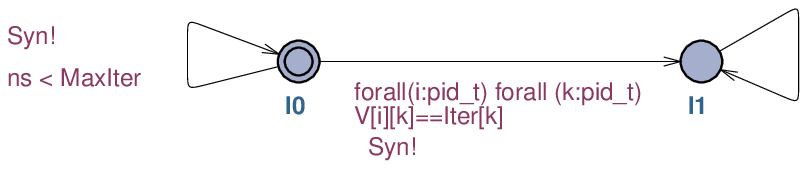}
\caption{Variant 1 of the concrete model: Process Controller}
\label{Preselection2}
\end{center}
\end{figure}

The first tentative to attenuate the state explosion problem is to
consider a variant of this model, where the selection of all
operations to be executed is performed at the beginning (before
executing the first operation). In this variant (see Figure
\ref{Preselection1}), each site begins with the selection of
signatures of its local operations. The variable $Compteur$, local
to the process \emph{Site}, is used to count the number of local
selected operations. The selection of local operations is done
synchronously with other sites to avoid all interleaving
executions resulting from the different selection orders of
operations by all sites. So, to achieve  this synchronization, we
have added another process called $Controller$ (see Figure
\ref{Preselection2}). Process $Controller$ uses the broadcast
channel $Syn$ to invite each site to choose an operation. This
synchronization allows to reduce the number of steps needed to
select all operations to be performed and avoids to consider
different numbering of operations. Indeed, the number of steps
passes from $MaxIter$ to $\underset{i \in
[0,NbSites-1]}{Max}(Iter[i])$. The number of creation orders
passes from $MaxIter!$ to $1$. The owner and signature of each
selected operation are stored in array \emph{Operations}. Each
operation is identified by its entry in
array \emph{Operations}.

After selecting all operations, each user executes the local and
non local operations in the same manner as in the previous model,
except that the synchronization on termination includes the
process $Controller$ which becomes the sender of signals $Syn$.

Results obtained for this model are reported in Table
\ref{TablePreselection}. We report, in column 5, the gain in both
space and time relatively to the concrete model, in the
form of ratios.\\
In UPPAAL, the definition of the system in this case is:\\
Sites(const pid\_t pid) = Site(pid);\\
\noindent \emph{Sites(const pid\_t pid) = Site(pid);}\\
\noindent \emph{system \ Sites, Controller; }
\begin{table}[h!]
\begin{center}
\caption{Variant 1 of the concrete model} \vspace{2pt}
\label{TablePreselection}
\begin{tabular}{|c||c||c||c||c|}
\hline $Alg. \ NbSites \ MaxIter$ & Prop. &Val.& Expl. / Comp. /
Time (s) & Gain
\\ \hline
 $Ellis \ 3 \ 3 $ & $\phi_1$  & true & 272665 / 349815 /  26.15  & 3.03 / 5.26 / 4.64\\
$Ellis \ 3 \ 3 $   & $\phi_2$ & true & 427494 / 427494 / 71.79 &
? \\\hline
$Ressel \ 3 \ 3$& $\phi_1$  & true & 277512 / 352740 / 26.40 & 3.00 / 5.25 / 4.63\\
$Ressel \ 3 \ 3$ & $\phi_2$ & true & 426188 / 426188 / 72.09 & ?  \\
   \hline
$Sun \ 3 \ 3$& $\phi_1$  & true & 43897 / 100612 / 4.03 & 19.06 / 18.86 / 30.35 \\
$Sun \ 3 \ 3$ & $\phi_2$ & true & 656102 / 656102 / 104.70 & ? \\
\hline $Suleiman \ 3 \ 3$& $\phi_1$  &false& 425252 / 425252 /
37.37 &
8.78 / 8.78 / 9.77 \\
$Suleiman \ 3 \ 3$ & $\phi_2$ & true & 425252 / 425252 / 72.63 & ?\\
$Suleiman \ 3 \ 4$ & $\phi_1$  & ? &  ? & ? \\
\hline $Imine \ 3 \ 3$ & $\phi_1$  & false&425252 / 425252 / 37.03 &  8.78 / 8.78 / 9.75\\
$Imine \ 3 \ 3$ & $\phi_2$ &true & 425252 / 425252 / 71.80 & ? \\
$Imine \ 3 \ 4$ & $\phi_1$  & ? & ? & ?\\
\hline
\end{tabular}
\end{center}
\end{table}

\subsection{Covering steps}

\begin{figure}[h]
\begin{center}
\includegraphics[width=0.569\textwidth]{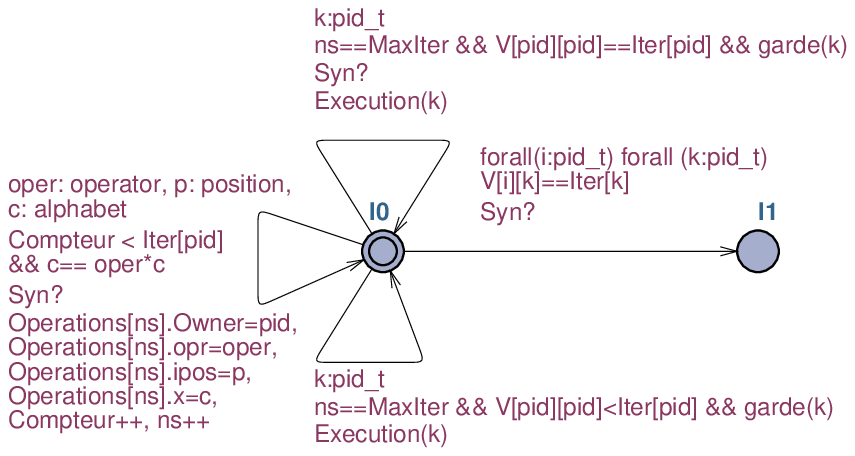}
\caption{Variant 2 of the concrete model: Process Site}
\label{Reduction1}
\includegraphics[width=0.45\textwidth]{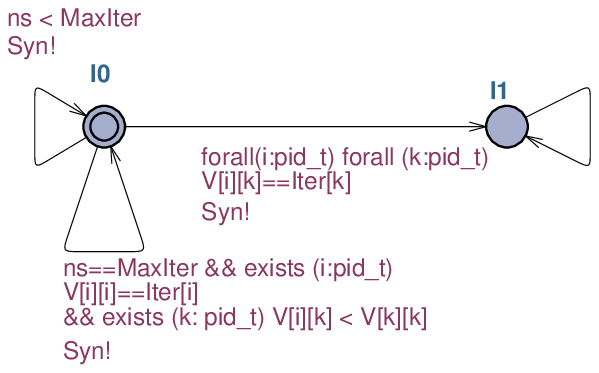}
\caption{Variant 2 of the concrete model: Process Controller}
\label{Reduction2}
\end{center}
\end{figure}

The changes proposed in the previous model allow a significant
gain in space and time. They are however not enough to achieve our
goal for some IT algorithms. To make more reductions,
we propose to group, in one step, the execution of non local
operations in sites which have finished the execution of their
local operations (see Figures \ref{Reduction1} and
\ref{Reduction2}).

This reduction preserves the convergence
property since when a site completes the execution of all local
operations, it does not send any information to other sites and
the execution of non local operations affects only the state of
the site. This agglomeration of steps is realized by means of the
broadcast channel $Syn$. When the process $Controller$ detects
that there is at least a site which has completed the execution of
all its local operations, it uses the channel $Syn$ to invite such
sites to execute synchronously one non local operation.

The results obtained, in this case, are reported in Table
\ref{TableSteps}. We give, in column 5, the gain in both space and
time relatively to the concrete model with preselecting of
operation signatures, in the form of ratios.

\begin{table}[h]
\begin{center}
\caption{Variant 2 of the concrete model} \label{TableSteps}
\vspace{2pt}
\begin{tabular}{|c||c||c||c||c|}
\hline $Alg.$ & Prop. & Val. & Expl. / Comp. / Time (s) & Gain
\\ \hline
 $Ellis \ 3 \ 3$ & $\phi_1$  & true &111700 / 240149 / 12.20  &  2.44 / 1.46 / 2.14 \\
 $Ellis \ 3 \ 3$ & $\phi_2$ & true &396569 / 396569 / 89.25 & 1.08 / 1.08 / 0.80\\ \hline
$Ressel \ 3 \ 3$ & $\phi_1$  & true & 120326 / 249233 / 13.21 & 2.31 / 1.42 / 2\\
 $Ressel \ 3 \ 3$ & $\phi_2$ & true & 395693 / 395693 / 89.99 & 1.08 / 1.08 / 0.80 \\ \hline
$Sun \ 3 \ 3$ & $\phi_1$  & true &  13513   /  43903 / 1.44 & 3.25 / 2.30 / 2.80\\
$Sun \ 3 \ 3$ & $\phi_2$ & true & 509431 / 509431 / 103.98 & 1.29
/ 1.29 / 1.01 \\ \hline
$Suleiman \ 3 \ 3$ & $\phi_1$  &false & 394877 / 394877 / 41.62 & 1.08 / 1.08 / 0.90\\
$Suleiman \ 3 \ 3$ & $\phi_2$ & true & 394877 / 394877 / 90.65 & 1.08 / 1.08 / 0.80 \\
$Suleiman \ 3 \ 4$ & $\phi_1$  & ? & ? & ?\\
\hline $Imine \ 3 \ 3$ & $\phi_1$  & false& 394877 / 394877 / 41.56 & 1.08 / 1.08 / 0.89\\
$Imine \ 3 \ 3$ & $\phi_2$ &true & 394877 / 394877 / 83.35 & 1.08 / 1.08 / 0.86 \\
$Imine \ 3 \ 4$& $\phi_1$  & ? & ? & ? \\
\hline
\end{tabular}
\end{center}
\end{table}

%% \begin{table}
%% \begin{center}
%% \caption{Concrete model with preselecting of operation signatures
%% and covering steps} \label{TableSteps} \vspace{2pt}
%% \begin{tabular}{|c||c||c||c||c|}
%% \hline $Alg.$ & Prop. & Val. & Expl. / Comp. / Time (s) & Gain
%% \\ \hline
%%  $Ellis$ & $\phi_1$  & true &111700 / 240149 / 12.20  &  2.44 / 1.46 / 2.14 \\

%%  $3 \ sites$ & $\phi_2$ & true &396569 / 396569 / 89.25 & 1.08 / 1.08 / 0.80\\ \hline
%% $Ressel$ & $\phi_1$  & true & 120326 / 249233 / 13.21 & 2.31 / 1.42 / 2\\
%%   $3 \ sites$ & $\phi_2$ & true & 395693 / 395693 / 89.99 & 1.08 / 1.08 / 0.80 \\ \hline
%% $Sun$ & $\phi_1$  & true &  13513   /  43903 / 1.44 & 3.25 / 2.30 / 2.80\\
%% $3 \ sites$ & $\phi_2$ & true & 509431 / 509431 / 103.98 & 1.29 /
%% 1.29 / 1.01 \\ \hline $Suleiman$ & $\phi_1$  &false & 394877 / 394877 / 41.62 & 1.08 / 1.08 / 0.90\\
%% $3 \ sites$ & $\phi_2$ & true & 394877 / 394877 / 90.65 & 1.08 / 1.08 / 0.80 \\
%% $4 \ sites$ & $\phi_1$  & ? & ? & ?\\
%% \hline $Imine$ & $\phi_1$  & false& 394877 / 394877 / 41.56 & 1.08 / 1.08 / 0.89\\
%% $3 \ sites$ & $\phi_2$ &true & 394877 / 394877 / 83.35 & 1.08 / 1.08 / 0.86 \\
%% $4 \ sites$ & $\phi_1$  & ? & ? & ? \\
%% \hline
%% \end{tabular}
%% \end{center}
%% \end{table}

In spite of these reductions, this model is still however
suffering from the state explosion problem. We have not been able
to verify the properties for $4$ sites. The verification was
aborted for a lack of memory when the number of states exceeds 4
millions. This state explosion problem is accentuated by the
number of operation signatures.
Indeed, the set of operation signatures is given by the following cartesian product:\\
$$((\{Del\} \times [0,L-1]) \cup (\{Ins\} \times [0,L-1] \times
A))^{MaxIter}$$ Its size increases exponentially with the number
of operations: $$ (L + (|A| \times L))^{MaxIter} = ((|A|+1) \times
L)^{MaxIter}$$

For example, the number of operation signatures is $5832$ for
$MaxIter=3$, $L=2 \times MaxIter$ and $|A|=2$. It reaches $331776$
for $MaxIter=4$ and $24300000$ for $MaxIter=5$. We propose, in the
following, another model where the instantiation of operation
signatures  is encapsulated in a function executed when the
construction of traces of all sites is completed.

\section{Symbolic model}
This model differs from the concrete model by the fact that the
instantiation of operation signatures is delayed until the
construction of execution traces of all sites is completed. So, it
has the same input data and the same set of processes.

\subsection{Variables}

The symbolic model uses the following variables (see Table
\ref{decl2}):
\begin{enumerate}
    \item The timestamp vectors of different sites ($V[NbSites][NbSites]$).
    \item Vector $Operations[MaxIter]$ to store the owner and the timestamp vector of each operation.
    \item Vectors $Trace[NbSites][MaxIter]$ to save the symbolic execution traces of
sites (the execution order of operations).
    \item Boolean variable $Detected$ to recuperate  the truth value of the convergence
    property.
    \item Vector $Signatures[MaxIter]$ to get back signatures
    $(operator, position, character)$ of operations which violate the convergence
    property.
    \item $List[2][MaxIter]$ to save operation signatures as they are
exactly executed in two sites. Recall that before executing a non
local operation, a site may transform it, using some
IT algorithm. Array $List$ is optional and used to
get back a counterexample which violates the convergence property
(exact traces). \item The broadcast channel $Syn$
 \end{enumerate}
For the same reasons as for the concrete model, all the above
variables are defined as global.

\begin{table}[h!]
\begin{center}
\caption{Declaration of constants, types and variables of the
symbolic model} \label{decl2}
\begin{tabular} {|l|}
\hline
// Declaration of constants \\
\emph{const  int  NbSites = 3};\\
\emph{const  int Iter[NbSites]= \{1,1,1\}}; \\
\emph{const int MaxIter=Iter[0]+Iter[1]+Iter[2]};\\
\emph{const int  L= 2*MaxIter};\\
\emph{const int  Del= 0};\\
\emph{const int  Ins= 0};\\
\emph{const int  Ellis= 0};\\
\emph{const int  Ressel= 1};\\
\emph{const int  Sun= 2};\\
\emph{const int  Suleiman= 3};\\
\emph{const int  Imine= 4};\\
\emph{const int[Ellis,Imine]  algo = Ellis};\\

// Declaration of types\\
\noindent \emph{typedef int[0, NbSites-1] pid\_t;}\\
\emph{typedef int[0, 1] alphabet;}\\
\emph{typedef int[Del, Ins] operator;}\\
\emph{typedef struct \{pid\_t Owner; int V[NbSites];\} operation\_t;}\\
\emph{typedef struct \{operator opr; alphabet x;
int pos;\} signature\_t;}\\
%\emph{typedef struct \{int numOp; int posC; \} trace\_t ;}\\
\emph{typedef struct \{int numOp; int posC; int a[MaxIter-1]; int b[MaxIter-1]; \} trace\_t ;}\\

// Declaration of variables\\
\emph{int[0,MaxIter-1] V[NbSites][MaxIter];}\\
\emph{operation\_t Operations[MaxIter];}\\
\emph{int[0,MaxIter-1] Trace[NbSites][MaxIter];}\\
\emph{signature\_t Signatures[MaxIter];}\\
\emph{trace\_t List[2][MaxIter];}\\
\emph{bool Detected =false;}\\
\emph{int[0, MaxIter] ns = 0;}\\

// Declaration of a broadcast channel\\
\noindent \emph{broadcast channel Syn;}\\
\hline
\end{tabular}
\end{center}
\end{table}

\subsection{Behavior of each site}

\begin{figure}[h!]
\begin{center}
\includegraphics[width=0.70\textwidth]{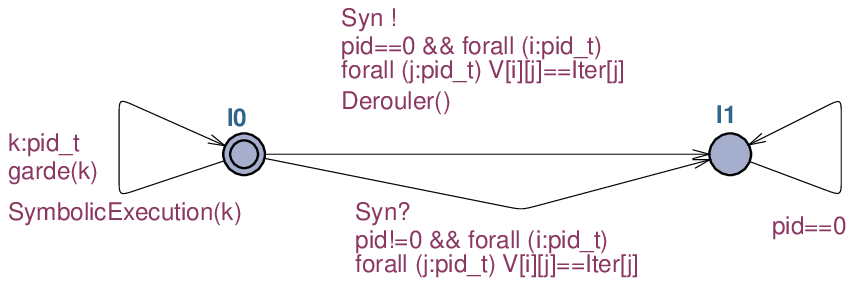}
\caption{The symbolic model} \label{Symbolic}
\end{center}
\end{figure}

As in the concrete model, behaviors of sites are similar and their
process template is shown in Figure \ref{Symbolic}. The only
parameter of the template is also the site identifier named $pid$.

\subsubsection{Symbolic operations and Traces}
Each site executes \emph{symbolically}, one by one, its local and
non local operations w.r.t. the causality principle. Operations
generated by each site are initially symbolic in the sense only
their owners and timestamp vectors are fixed and stored in array
$Operations$. As in the previous models, each operation has its
own identifier corresponding to its entry in array $Operations$.
The execution order of symbolic operations (symbolic traces) are
got back in array $Trace$. In $Trace[i][n]$, we get back the
identifier of the $n^{th}$ operation
executed by site $i$.

The signature of each operation is instantiated when the execution
traces of all sites are completed.  Vectors $Signatures$ and
$List$ are used to get back operation signatures and concrete
execution traces which violate the convergence property (i.e. a counterexample).

\subsubsection{Symbolic execution of local operations}
As for the concrete model, a local operation can be executed by
site $pid$ if there is at least a local operation not yet executed
(i.e. $V[pid][pid] < Iter[pid]$). When an operation is executed
locally, its timestamp vector is set to the timestamp vector of
its site. The owner and the timestamp vector of the operation are
stored in $Operations$. Its entry in $Operations$ is stored in
$Trace[pid]$. Its broadcast to other sites is also simulated by
incrementing the number of local operations executed
($V[pid][pid]++$) (see functions $garde$ and $SymbolicExecution$
in the Appendix).

\subsubsection{Symbolic execution of non local operations}
The condition to be satisfied to execute a non local operation is
the same as the one of the concrete model. %A site $pid$ can
%execute a non local operation, if the timestamp vector of the
%operation
% is less or equal to the timestamp vector of the site $pid$ ($\forall j \in [0,NbSites-1],
%V[pid][j] >= Operations[num].V[j]$, $num$ being the operation
%identifier).\\
Recall that, the transformation and effective execution of
operations (Insert and Delete) are not performed at this level.
They are realized when the
construction of all traces is completed.

\subsubsection{Effective execution of operations} When all
sites complete the construction of their respective  traces, they
are forced to perform synchronously, via the broadcast channel
$Syn$, edges connecting locations $l0$ and $l1$ of all sites. The
action of edge connecting locations $l0$ and $l1$ of site $0$ is
devoted to testing all signatures possibilities of operations and
then verifying the convergence property. The test of all these
possibilities is encapsulated in a C-function, called $Derouler$
which is stopped as soon as the violation of the convergence
property is detected. In this case, signatures of operations and
exact traces of two sites which violate the convergence property
are returned in vectors $Signatures$ and $List$, and the variable
$Detected$ is set to $true$.

\subsubsection{Verification of properties}

To verify whether the convergence property is satisfied or not, it
suffices to use the variable $Detected$. This variable is set to
$true$ when the convergence propriety is violated. So, UPPAAL's
$CTL$ formula $E\Diamond \ Detected$ is satisfied if and only if the
convergence propriety is violated.

We have tested the IT algorithms considered here using the
symbolic model. We report in Table \ref{TableSymbolic} the results
obtained, for two properties: absence of deadlocks ($\phi_2:$
$A\Box \ not \ deadlock$) and the violation of the convergence
property ($\phi_1': E\Diamond \ Detected$), in case of 3 and 4
sites, 3 and 4 operations, and a window of text of length
$2*MaxIter$.

\begin{table}[h!]
\begin{center}
\caption{Symbolic model} \label{TableSymbolic} \vspace{2pt}
\begin{tabular}{|c||c||c||c||c|}
\hline $Alg. \ NbSites \ MaxIter$ & Prop. &Val.& Expl. / Comp. /
Time (s) & Gain
\\ \hline
 $Ellis \ 3 \ 3$  & $\phi_1'$  & true & 1625 / 1739 /  0.14 & 68.74 /138.10 / 87.14 \\
$Ellis \ 3 \ 3$  & $\phi_2$ & true & 1837 / 1837 / 0.68 & 215.88 / 215.88 / 131.25 \\
\hline
$Ressel \ 3 \ 3$ & $\phi_1'$  & true & 1637 / 1751 / 0.25 & 73.50 / 142.34 / 52.84\\
  $Ressel \ 3 \ 3$ & $\phi_2$ & true & 1837 / 1837 / 1.63 & 215.88 / 215.88 / 131.25\\ \hline
$Sun \ 3 \ 3$ & $\phi_1'$  & true &  1625 / 1739 / 0.14 &  8.32 / 25.25 / 10.29 \\
$Sun \ 3 \ 3$ & $\phi_2$ & true & 1837 / 1837 / 0.38 &  277.32 /
277.32 / 273.63 \\ \hline
$Suleiman \ 3 \ 3$ & $\phi_1'$  &false& 1837 / 1837 / 0.83 &  214.96 / 214.96 / 40.84 \\
$Suleiman \ 3 \ 3$& $\phi_2$ & true &  1837 / 1837 / 2.22 &  214.96 / 214.96 / 40.84\\
$Suleiman \ 3 \ 4$ & $\phi_1'$  & true & 18450 / 19380 /  2.45  & ? \\
\hline
$Imine \ 3 \ 3$ & $\phi_1'$  & false& 1837 / 1837 / 0.81 & 214.96 / 214.96 / 40.84\\
$Imine \ 3 \ 3$ & $\phi_2$ &true &  1837  / 1837 / 2.18 &  214.96 / 214.96 / 40.84\\
 $Imine \ 3 \ 4$ & $\phi_1'$  & true & 18401 /  19331 /
2.45 & ?
\\ \hline
\end{tabular}
\end{center}
\end{table}

%% \begin{table}
%% \begin{center}
%% \caption{Case of the symbolic model} \label{TableSymbolic}
%% \vspace{2pt}
%% \begin{tabular}{|c||c||c||c||c|}
%% \hline $Alg.$ & Prop. &Val.& Expl. / Comp. / Time (s) & Gain
%% \\ \hline
%%  $Ellis$  & $\phi_1'$  & true & 1625 / 1739 /  0.14 & 68.74 /138.10 / 87.14 \\
%%  $3 \ sites$ & $\phi_2$ & true & 1837 / 1837 / 0.68 & 215.88 / 215.88 / 131.25 \\
%% \hline
%% $Ressel$ & $\phi_1'$  & true & 1637 / 1751 / 0.25 & 73.50 / 142.34 / 52.84\\
%%   $3 \ sites$ & $\phi_2$ & true & 1837 / 1837 / 1.63 & 215.88 / 215.88 / 131.25\\ \hline
%% $Sun$ & $\phi_1'$  & true &  1625 / 1739 / 0.14 &  8.32 / 25.25 / 10.29 \\
%% $3 \ sites$ & $\phi_2$ & true & 1837 / 1837 / 0.38 &  277.32 /
%% 277.32 / 273.63 \\ \hline
%% $Suleiman$ & $\phi_1'$  &false& 1837 / 1837 / 0.83 &  214.96 / 214.96 / 40.84 \\
%% $3 \ sites$ & $\phi_2$ & true &  1837 / 1837 / 2.22 &  214.96 / 214.96 / 40.84\\
%% $4 \ sites$ & $\phi_1'$  & true & 3413889 / 3413889 /  493.25  & ? \\
%% \hline
%% $Imine$ & $\phi_1'$  & false& 1837 / 1837 / 0.81 & 214.96 / 214.96 / 40.84\\
%% $3 \ sites$ & $\phi_2$ &true &  1837  / 1837 / 2.18 &  214.96 / 214.96 / 40.84\\
%%  $4 \ sites$ & $\phi_1'$  & true & 3413889 /  3413889 /
%% 490.05 & ?
%% \\ \hline
%% \end{tabular}
%% \end{center}
%% \end{table}
%
\subsection{Pre-numbering symbolic operations and covering steps}
We have tested the effect of the pre-numbering of symbolic
operations and covering steps on the symbolic model. The
pre-numbering of symbolic operations is somewhat a symbolic
version of the concrete model with preselecting of operation
signatures. Results obtained, in this case, are reported in Table
\ref{TablePreselctionReductionSymbolicbis}.

\begin{table}[h!]
\begin{center}
\caption{Variant 1 of the symbolic model}
\label{TablePreselctionReductionSymbolicbis} \vspace{2pt}
\begin{tabular}{|c||c||c||c|}
\hline $Alg. \ NbSites \ MaxIter$ & Prop. &Val.& Expl. / Comp. / Time (s)  \\
\hline
 $Ellis \ 3 \ 3$ & $\phi_1'$  & true & 640 / 899 / 0.14 \\
 $Ellis \ 3 \ 3$ & $\phi_2$ & true & 1267 / 1267 / 1.76\\ \hline
$Ressel \ 3 \ 3$ & $\phi_1'$  & true & 680 / 935 / 0.30 \\
$Ressel \ 3 \ 3$ & $\phi_2$ & true &  1267 / 1267 / 5.33  \\\hline
$Sun \ 3 \ 3$ & $\phi_1'$  & true &  640 / 899 / 0.16 \\
$Sun \ 3 \ 3$ & $\phi_2$ & true & 1267 / 1267 / 1.01  \\
\hline $Suleiman \ 3 \ 3$ & $\phi_1'$  & false & 1267 / 1267 / 2.35 \\
$Suleiman \ 3 \ 3$& $\phi_2$ & true &  1267 / 1267 / 6.73  \\
$Suleiman \ 3 \ 4$& $\phi_1'$  & true & 643 / 1110 /  0.24 \\
\hline
$Imine \ 3 \ 3$ & $\phi_1'$  & false & 1267 / 1267 / 2.30\\
$Imine \ 3 \ 3$ & $\phi_2$ & true &  1267  / 1267 / 6.61 \\
$Imine \ 3 \ 4$ & $\phi_1'$  & true & 643 / 1110 / 0.33 \\
\hline
\end{tabular}
\end{center}
\end{table}

\begin{figure}[h!]
\begin{center}
\includegraphics[width=0.60\textwidth]{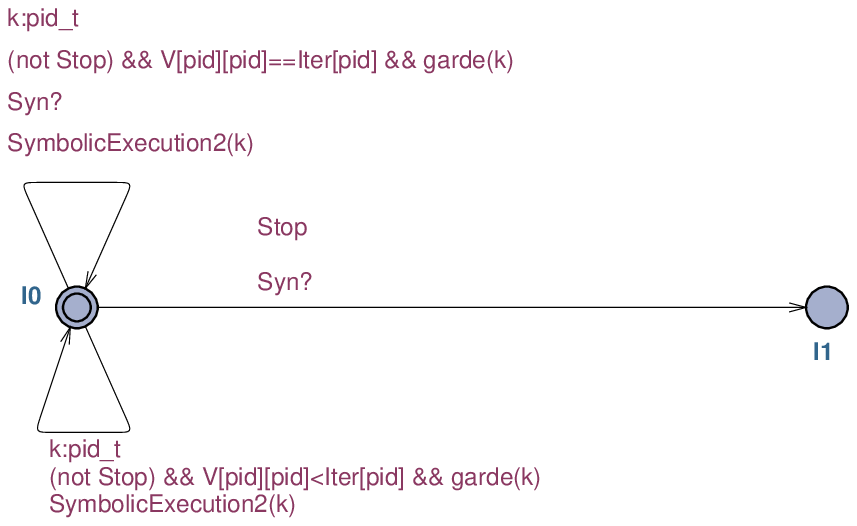}
\caption{Variant 1 of the symbolic model: Process Site}
\label{PreselectionReductionSymbolic1}
\includegraphics[width=0.67\textwidth]{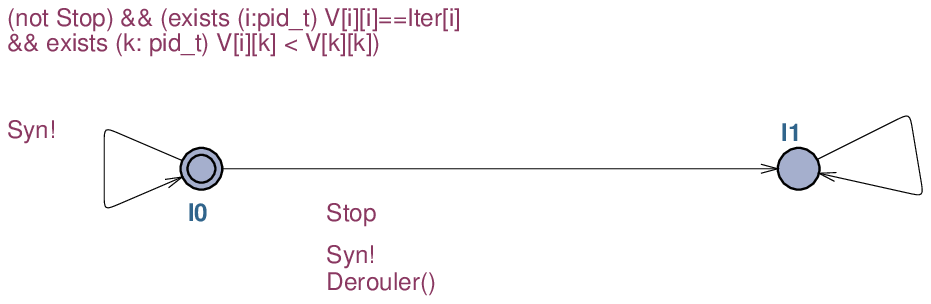}
\caption{Variant 1 of the symbolic model: Process Controller}
\label{PreselectionReductionSymbolic2}
\end{center}
\end{figure}

We have also considered a variant of the symbolic model where, in
addition to the previous changes, we force to stop the
construction of symbolic traces of other sites as soon as two
sites have completed their own traces (see Figures
\ref{PreselectionReductionSymbolic1} and
\ref{PreselectionReductionSymbolic2}). The boolean variable $Stop$
is set to $true$ in function $SymbolicExecution2(k)$ as soon
as two any sites complete the symbolic execution of all operations.

As sites have symmetrical behaviors, this restriction does not
alter the convergence property. Results obtained, in this case,
are reported in Table \ref{TableSymbolicEarlyStop}.

\begin{table}[h!]
\begin{center}
\caption{Variant 2 of the symbolic model}
\label{TableSymbolicEarlyStop}\vspace{2pt}
\begin{tabular}{|c||c||c||c|}
\hline $Alg. \ NbSites \ MaxIter$ & Prop. &Val.& Expl. / Comp. / Time (s) \\
\hline
 $Ellis \ 3 \ 3$ & $\phi_1'$  & true & 95 / 157 / 0.07   \\
 $Ellis \ 3 \ 3$ & $\phi_2$ & true & 278 / 278 / 0.83 \\ \hline
$Ressel \ 3 \ 3$ & $\phi_1'$  & true & 95 / 157 / 0.09  \\
  $Ressel \ 3 \ 3$  & $\phi_2$ & true &  278 / 278 / 1.00   \\
 \hline
$Sun \ 3 \ 3$ & $\phi_1'$  & true &  95 / 157 / 0.08 \\
$Sun \ 3 \ 3$ & $\phi_2$ & true &  278 / 278 / 0.70   \\\hline
$Suleiman \ 3 \ 3$ & $\phi_1'$  & false & 278 / 278 / 0.59    \\
$Suleiman \ 3 \ 3$ & $\phi_2$ & true &  278 / 278 / 1.63   \\
$Suleiman \ 3 \ 4$ & $\phi_1'$  & true & 643 / 1125 / 7.79   \\
\hline
$Imine \ 3 \ 3$ & $\phi_1'$  & false & 278 / 278 / 0.59  \\
$Imine \ 3 \ 3$& $\phi_2$ & true &  278  / 278 / 1.58 \\
$Imine \ 3 \ 4$ & $\phi_1'$  & true & 643 / 1125 / 7.36 \\
\hline
\end{tabular}
\end{center}
\end{table}

%% \begin{table}
%% \begin{center}
%% \caption{Applying the covering steps, pre-numbering of symbolic
%% operations and early stop to the symbolic model}
%% \label{TableSymbolicEarlyStop}\vspace{2pt}
%% \begin{tabular}{|c||c||c||c|}
%% \hline $Alg.$ & Prop. &Val.& Expl. / Comp. / Time (s) \\
%% \hline
%%  $Ellis$ & $\phi_1'$  & true & 95 / 157 / 0.07   \\
%%  $3 \ sites$ & $\phi_2$ & true & 278 / 278 / 0.83 \\ \hline
%% $Ressel$ & $\phi_1'$  & true & 95 / 157 / 0.09  \\
%%   $3 \ sites$ & $\phi_2$ & true &  278 / 278 / 1.00   \\
%%  \hline
%% $Sun$ & $\phi_1'$  & true &  95 / 157 / 0.08 \\
%% $3 \ sites$ & $\phi_2$ & true &  278 / 278 / 0.70   \\\hline
%% $Suleiman$ & $\phi_1'$  & false & 278 / 278 / 0.59    \\
%% $3 \ sites$ & $\phi_2$ & true &  278 / 278 / 1.63   \\
%% $4 \ sites$ & $\phi_1'$  & true & 14939 / 27598 /  2658.57   \\
%% \hline
%% $Imine$ & $\phi_1'$  & false & 278 / 278 / 0.59  \\
%% $3 \ sites$& $\phi_2$ & true &  278  / 278 / 1.58 \\
%% $4 \ sites$ & $\phi_1'$  & true & 14939 / 27598 / 2347.36 \\
%% \hline
%% \end{tabular}
%% \end{center}
%% \end{table}
%
\subsection{Symbolic model without timestamp vectors}

Another factor which contributes to the state explosion problem is
the timestamp vectors of different sites and operations. These
vectors are used to ensure the causality principle.  We have
$NbSites+MaxIter$ timestamp vectors (one per site and one per
operation). Each timestamp vector consists of $NbSites$ elements.
The range of each element $i$ is $0..Iter[i]-1$. To attenuate the
state explosion problem caused by timestamp vectors, we propose,
in the following model, to replace these timestamp vectors with a
relation of dependence over operations and the vector
$CO[NbSites]$ which indicates for each site the number of
operations yet executed by the site till now.

This model offers the possibility to fix a dependence relation
over operations and to test whether an IT algorithm
works or not under some relation of dependence. %Indeed, we have been able to show
%that \emph{Suleiman}'s algorithm does satisfy the convergence
%property for 4 concurrent operations. The convergence property is
%however satisfied for \emph{Imine}'s algorithm for all tested examples.\\

\begin{figure}[h!]
\begin{center}
\includegraphics[width=0.60\textwidth]{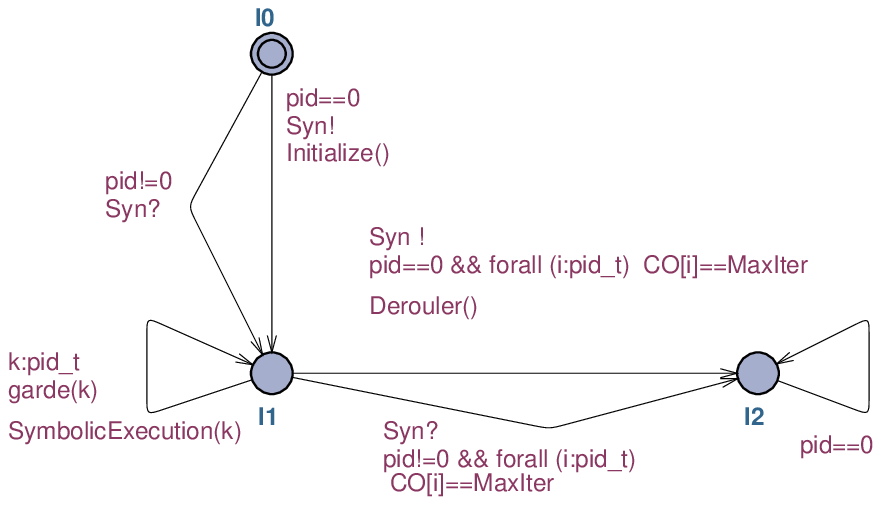}
\caption{Variant 3 of the symbolic model}
\label{SymbolicWithoutVectors}
\end{center}
\end{figure}

This variant of the symbolic model is shown in Figure
\ref{SymbolicWithoutVectors}. This model consists of $NbSites$
processes $Site$. These sites start by executing together edges
connecting locations $l0$ and  $l1$ (initialization phase). This
phase (function $Initialize$), performed by site $0$, consists of
setting input data (initial text, dependent operations...). All
sites remain in locations $l1$ until they finish the execution of
all operations. Then, they synchronize on termination to reach
together their respective locations $l2$.

In Table \ref{TablePreselctionReductionSymbolic}, we report
results obtained for the case of independent operations and the
case of two dependent operations. We have also considered a
variant of this model where the construction of symbolic traces
are stopped as soon as two sites have completed the execution of
all operations. Results obtained for this variant are reported in
Table \ref{TablePreselctionEarlyStopSymbolic}.

\begin{table}[h!]
\begin{center}
\caption{Variant 3 of the symbolic model}
\label{TablePreselctionReductionSymbolic} \vspace{2pt}
\begin{tabular}{|c||c||c||c|}
\hline $Alg. \ NbSites \ MaxIter$ & Prop. &Val.& Expl. / Comp. / Time (s) \\
\hline
 $Ellis \ 3 \ 3$ & $\phi_1'$  & true & 135 / 143 / 0.15  \\
 $Ellis \ 3 \ 3$   & $\phi_2$ & true & 150 / 150 / 1.11 \\\hline
$Ressel \ 3 \ 3$ & $\phi_1'$  & true & 135 / 143 / 0.08 \\
$Ressel \ 3 \ 3$  & $\phi_2$ & true &  150 / 150 / 0.46  \\ \hline
$Sun \ 3 \ 3$ & $\phi_1'$  & true &  135 / 143 / 0.09 \\
$Sun \ 3 \ 3$ & $\phi_2$ & true &  150 / 150 / 0.18 \\
\hline
$Suleiman \ 3 \ 3$ & $\phi_1'$  & false & 150 / 150 / 0.44   \\
$Suleiman \ 3 \ 3$ & $\phi_2$ & true &  150 / 150 / 1.11 \\
$Suleiman \ 3 \ 4 \ 0\rightarrow 1$ & $\phi_1'$  & true & 439 /457  / 0.53  \\
$Suleiman \ 4 \ 4$ & $\phi_1'$  & false & 67362 / 67362 /  1045.78  \\

\hline
$Imine \ 3 \ 3$ & $\phi_1'$  & false & 150 / 150 / 0.42  \\
$Imine \ 3 \ 3$& $\phi_2$ & true &  150  / 150 / 1.12 \\
$Imine \ 3 \ 4 \ 0\rightarrow 1$ & $\phi_1'$  & true & 439 /457  / 0.26 \\
$Imine \ 4 \ 4$ & $\phi_1'$  & false & 67362 / 67362 / 981.02   \\
\hline
\end{tabular}
\end{center}
\end{table}

\begin{table}[h!]
\begin{center}
\caption{Variant 4 of the symbolic model}
\label{TablePreselctionEarlyStopSymbolic} \vspace{2pt}
\begin{tabular}{|c||c||c||c|}
\hline $Alg. \ NbSites \ MaxIter$ & Prop. &Val.& Expl. / Comp. / Time (s) \\
\hline
 $Ellis \ 3 \ 3$ & $\phi_1'$  & true &  18/ 26 / 0.10  \\
  $Ellis \ 3 \ 3$   & $\phi_2$ & true & 33 / 33 /0.32  \\\hline
$Ressel \ 3 \ 3$ & $\phi_1'$  & true &  18 / 26 / 0.11 \\
  $Ressel \ 3 \ 3$ & $\phi_2$ & true &  33 / 33 /   0.47\\ \hline
$Sun \ 3 \ 3$ & $\phi_1'$  & true & 18  / 26 / 0.11 \\
$Sun \ 3 \ 3$ & $\phi_2$ & true &  33 / 33 / 0.16 \\
\hline
$Suleiman \ 3 \ 3$ & $\phi_1'$  & false & 33 / 33 / 0.45   \\
$Suleiman \ 3 \ 3$& $\phi_2$ & true &  33  /33  /  1.12\\
$Suleiman \ 3 \ 4 \ (0\rightarrow 1)$ & $\phi_1'$  & true &  40 / 58 / 0.2 \\
$Suleiman \ 4 \ 4 $ & $\phi_1'$  & false & 3986 / 3986 /  968.29   \\
\hline
$Imine \ 3 \ 3$ & $\phi_1'$  & false & 33/ 33 / 0.42 \\
$Imine \ 3 \ 3$ & $\phi_2$ & true &  33  /33  /  1.08\\
$Imine \ 3 \ 4 \ (0\rightarrow 1)$ & $\phi_1'$  & true & 40 / 58 / 0.2 \\
$Imine \ 4 \ 4$ & $\phi_1'$  & false & 3986 / 3986 /  967.08  \\
\hline
\end{tabular}
\end{center}
\end{table}

\section{Related Work}
To our best knowledge, there exists only one work on analyzing
OT algorithms~\cite{Imi06}. In this work, the
authors proposed a formal framework for modeling and verifying IT
algorithms with algebraic specifications. For checking the
IT properties $TP1$ and $TP2$, they used a theorem
prover based on advanced automated deduction techniques. This
theorem proving approach turned out very valuable because many
bugs have been detected in well-known IT algorithms, as shown in
Table~\ref{tab:ot_th}.

\begin{table}[h!]
\begin{center}
\caption{Bugs detected by theorem prover-based approach.}\label{tab:ot_th}
\begin{tabular}{|l|c|c|} \hline
\textbf{IT algorithms}       & $\mathbf{TP1}$      &
$\mathbf{TP2}$
\\ \hline Ellis et \textit{al.}        & violated & violated   \\
\hline Ressel et \textit{al.}       & violated   & violated   \\
\hline Sun et \textit{al.}          & violated   & violated   \\
\hline Suleiman et \textit{al.}     & correct    & violated   \\
\hline Imine et \textit{al.}        & correct    & violated   \\
\hline
\end{tabular}
%\caption{Bugs detected by theorem prover-based approach.}\label{tab:ot_th}
%\vspace{-5mm}
\end{center}
\end{table}

It is clear that the theorem prover-based approach is appropriate
to detect bugs which may lead to potential divergence situations.
Nevertheless, it is less efficient in many cases as it does not
give how to reach these bugs. In other terms, it is unable to
output a complete scenario leading to divergence situation. Note
that a scenario consists of:
\begin{inparaenum}[(i)]
\item a number of sites as well as operations generated on these
  sites;
\item execution orders which show how each site integrates all
operations.
\end{inparaenum}

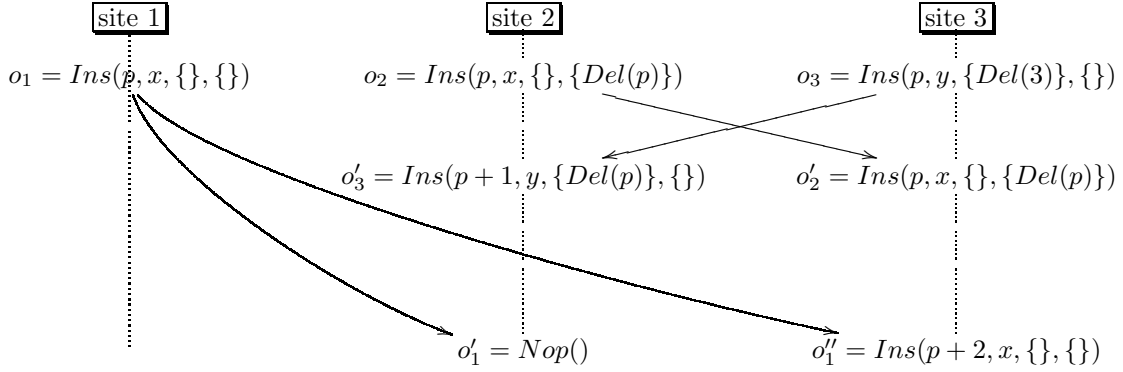
\begin{figure}[h!]
\centerline{\xymatrix@C=30pt@M=2pt@R=10pt{ *+[F-,]\txt{site 1}
\ar@{.}'[dd][ddddddd] & *+[F-,]\txt{site 2}
\ar@{.}'[d]'[dd]'[ddd]'[dddd]'[ddddd][ddddddd] &
*+[F-,]\txt{site 3} \ar@{.}'[d]'[dd]'[ddd]'[dddd]'[ddddd][ddddddd] \\
o_1 = Ins(p,x,\{\},\{\}) \ar@(d,l)[ddddddr]
\ar@(d,l)[ddddddrr]& o_2 = Ins(p,x,\{\},\{Del(p)\}) \ar[ddr]
&
o_3 = Ins(p,y,\{Del(3)\},\{\}) \ar[ddl] \\%|!{[l];[dd]}\hole \\
 &  &  \\
& o'_3 = Ins(p+1,y,\{Del(p)\},\{\})  & o'_2 = Ins(p,x,\{\},\{Del(p)\}) \\
 & &  \\
 & &  \\
 & &  \\
& o'_1 = Nop() & o''_1 = Ins(p+2,x,\{\},\{\}) \\
&  &   \\
}} \caption{$TP2$ violation for Suleiman's algorithm.}
\label{fig:suleimanko_small}
\end{figure}

It is important to find a scenario against a potential bug because
it enables us not only to get a concrete evidence that the
divergence situation exists, but also to have a better insight
into the shortcoming of IT algorithms. For example, consider the
IT algorithm proposed by Suleiman  et
\textit{al.}~\cite{suleiman97}. A theorem prover-based
verification revealed a $TP2$ violation in this
algorithm~\cite{Imine-PhD06}, as illustrated in
Figure~\ref{fig:suleimanko_small}. As this is related to $TP2$
property, there are three concurrent
operations (for all positions $p$ and all characters $x$ and $y$ such
that $Code(x) < Code (y)$):\\
$o_1 = Ins(p,x,\{\},\{\})$, $o_2 =
Ins(p,x,\{\},\{Del(p)\})$ and $o_3 =
Ins(p,y,\{Del(p)\},\{\})$ with the transformations
$o'_3=IT(o_3,o_2)$, $o'_2=IT(o_2,o_3)$,
$o'_1=IT^*(o_1,[o_2;o'_3])$ and $o''_1=IT^*(o_1,[o_3;o'_2])$.

However, the theorem prover's output gives no information
about  whether this $TP2$ violation is reachable or not. Indeed,
we do not know how to obtain $o_2$ and $o_3$ (their $av$ and $ap$
parameters are not empty respectively) as they are necessarily the
results of transformation against other operations that are not
given by the theorem prover.

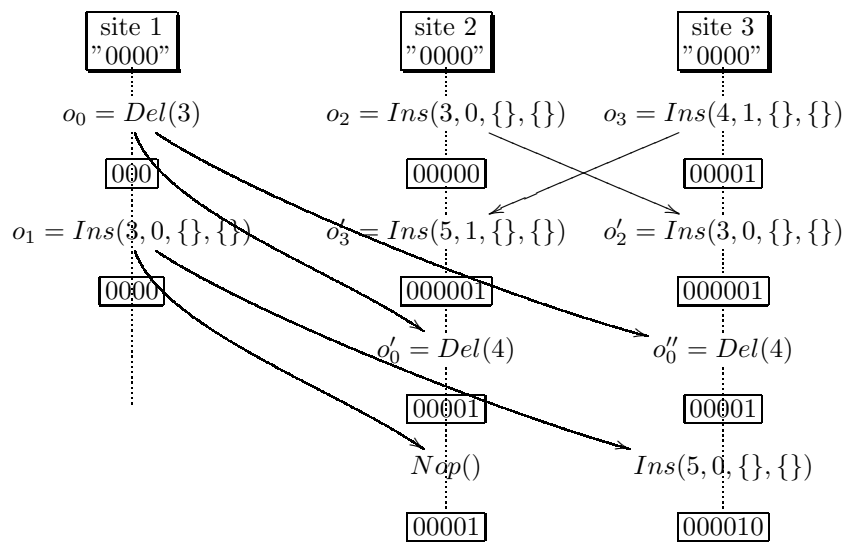
\begin{figure}[h!]
%\begin{tiny}
\centerline{\xymatrix@C=10pt@M=2pt@R=10pt{ *+[F-,]\txt{site 1 \\
"0000"} \ar@{.}'[d][dddddd] & & *+[F-,]\txt{site 2 \\ "0000"}
\ar@{.}'[d]'[dd]'[ddd]'[dddd]'[ddddd][dddddddd] &
*+[F-,]\txt{site 3 \\ "0000"} \ar@{.}'[d]'[dd]'[ddd]'[dddd]'[ddddd][dddddddd] \\
o_0 = Del(3) \ar@(d,ul)[ddddrr] \ar@(dr,l)[ddddrrr]& & o_2 =
Ins(3,0,\{\},\{\}) \ar[ddr] &
o_3 = Ins(4,1,\{\},\{\}) \ar[ddl] \\
*+[F]{000} & & *+[F]{00000} & *+[F]{00001} \\
o_1 = Ins(3,0,\{\},\{\}) \ar@(d,ul)[ddddrr]
\ar@(dr,l)[ddddrrr]& & o'_3 = Ins(5,1,\{\},\{\}) &
o'_2 = Ins(3,0,\{\},\{\}) \\%|!{[l];[dd]}\hole \\
*+[F]{0000} & & *+[F]{000001} & *+[F]{000001} \\
& & o'_0 = Del(4)  & o''_0 = Del(4) \\
& & *+[F]{00001} & *+[F]{00001} \\
& & Nop() & Ins(5,0,\{\},\{\}) \\
& & *+[F]{00001} & *+[F]{000010}  \\
}} \caption{Complete divergence scenario for Suleiman's
algorithm.} \label{fig:suleimanko}
%\end{tiny}
\end{figure}

Using our model-checking-based technique, we can get a complete
and informative scenario when a bug is detected. Indeed, the
output contains all necessary operations and the step-by-step
execution that lead to divergence situation. Thus, by
model-checking verification, the existence of the $TP2$ violation
depicted in Figure~\ref{fig:suleimanko_small} is proved (or
certified) by the scenario given in Figure~\ref{fig:suleimanko},
where $o_0$, $o_2$ and $o_3$ are pairwise concurrent and $o_0
\rightarrow o_1$.

As they are the basis cases of the convergence property, $TP1$ and
$TP2$ are sufficient to ensure the data convergence for any number
of concurrent operations which can be performed in any order.
Thus, a theorem prover-based approach remains better for proving
that some IT algorithm satisfies $TP1$ and $TP2$. But it is
partially automatable and, in the most cases, less informative
when divergence bugs are detected. A model-checking-based approach
is fully automatable for finding divergence scenarios.
Nevertheless, it is more limited as the convergence property can
be exhaustively evaluated on only a specific finite state space.

\section{Conclusion}
We proposed here a model-checking technique, based on formalisms
used in tool UPPAAL, to model the behavior of replication-based
distributed collaborative editing systems. To cope with the severe
state explosion problem of such systems, we exploited their
features and those of tool UPPAAL to establish and apply some
abstractions and reductions to the model. The verification of the
model and its variants have been performed with the model-checker
module of UPPAAL. An interesting and useful feature of this module
is to provide, in case of failure of the tested property, a trace
of an execution for which the property is not satisfied.  We used
this feature to give counterexamples for five IT algorithms,
proposed in the literature in order to ensure the convergence
property in the replication-based distributed distributed
collaborative editing systems. Using our model-checking technique
we found an upper bound for ensuring the data convergence in such
systems. Indeed, when the number of sites exceeds 2 the
convergence property is not achieved for all IT algorithms
considered here.

However, the serious drawback of the model-checking is the state
explosion. So, in future work, we plan to investigate the
following problems:
\begin{itemize}
\item It is interesting to find, under which conditions, the
  model-checking verification problem can be reduced to a finite-state
  problem.
\item Combining theorem-prover and model-checking approaches in
order
  to attenuate the severe state explosion problem.
\end{itemize}

\newpage

\bibliographystyle{abbrv}
\bibliography{mybib}

\newpage

\section{Appendix: codes of different Functions}

\noindent \textbf{Functions:} This appendix is devoted to the main
functions used in models proposed here for the replication-based
distributed groupware systems. These functions are defined as
local functions of process \emph{Site}. Therefore, they have as an
implicit parameter the process identifier of the process
\emph{Site}. We give here the code of the following functions:
\begin{enumerate}
    \item  Function $garde$ tests whether a site $pid$ can execute an
operation of some site $k$ (see Algorithm  \ref{algo_Garde}.
Recall that $pid$ is the parameter of the process Site and then an
implicit parameter of this function. A local operation (i.e., case
$k=pid$) can be executed if the number of local operations
executed till now does not reach the maximal number of local
operations to be executed (i.e., $V[pid][pid] < Iter[pid]$). A non
local operation (i.e., case $k \neq pid$) can be executed if it
satisfies the causality principle ($\forall j:pid_t, (V[pid][j]
\geq Operations[num].V[j])$, where $num$ is the operation
identifier). $garde \ (pid\_t \ k)$ used to test whether a site
$pid$ can execute an operation of site $k$ or not.

\item Function $Execution$ is devoted to manage the
      construction of concrete traces and execution of operations.
      It initializes copies of texts when it is
called for the first time, gets the identifier and the signature
of the operation to be executed. In case of a local operation, it
sets the timestamp vector of the operation to the one of the site.
Then, it calls the IT procedure and actualizes its
proper timestamp vector.
    \item Function $Operation$  implements different operation codes
(insert and delete). It executes an operation on the text copy of
the site $pid$. Operations with inconsistent signatures (i.e.,
parameter \emph{position} is outside the considered window of the
text) are ignored.

\item Functions $Transformation$, $TransformR$ and $ReOrder$ are
used to ensure the common treatment of the IT
algorithms.
\begin{enumerate}
 \item Function $Transformation$ lunches
the effective transformation process in case the operation is not
local (function $TransformR$). The resulting operation is executed
by calling function $Operation$.

\item Functions $TransformR$ is devoted to the integration process
of a non local operation $O$ (see section \ref{sec:integration}).
This process starts with reordering the operations executed till
now (history, operations of vector $List$). Therefore, it
transforms, using an IT algorithm, the resulting list and
operation $O$ relatively to all concurrent operations of the
reordered list of operations while dealing with the partial
concurrency problem (see section \ref{sec:partialconcurrency}.
Note that, tool $UPPAAL$ does not allow recursive functions. To
overcome this limitation, for implementation purpose, we have
rewritten this function.

\item Function $ReOrder$ reorders a list of operations $List$ in
order to put all operations dependant of an operation  $op$ on the
top. The resulting list is returned in $List1$.
\end{enumerate}
 \item Function $Concurrent$ tests whether two operations $op1$ and $op2$
are concurrent.

\item Function $SymbolicExecution$ is devoted to manage the
construction of symbolic traces. It is the same as function
$Execution$, except that the effective execution of an operation
is replaced by its insertion in the trace vector $Trace[pid]$ of
the site $pid$. This vector is used to get back the execution
order of operations in site $pid$.

\item $SymbolicExecution2$ is the same as $SymbolicExecution$
except that we force it termination as soon as any two sites have
completed their execution.
\end{enumerate}

\begin{algorithm}[h!]
 \caption{: Function garde} \label{algo_Garde}
\begin{algorithmic}
 \STATE $bool \ garde (pid\_t \ k)$
    \STATE $int \ i,j$;\\
    \COMMENT{ $pid$ is the parameter of the process Site}\\
    \COMMENT{ Each function of the process has the pid as an implicit
    parameter}
     \IF{ $(pid ==k)$} \STATE return $V[pid][pid] < Iter[pid]$; \ENDIF
    % \COM get the identifier of the $V[pid][k]$ nth operation of $k$\\
    \IF { ($V[pid][k] < V[k][k]$)}
            \FOR { (\ $i=0, \ j=0; \ i < MaxIter \ \&\& \ j < = V[pid][k];  \ i++$)}
                   \IF{ $(Operations[i].Owner \ == \ k)$ }
                    \STATE $j++$ ;  \ENDIF
            \ENDFOR
            % \COM $i-1$ is the identifier of the operation
              \FOR{ ($j=0; \ j<NbSites ; \ j++$)}
                 \IF{$(V[pid][j] < Operations[i-1].V[j])$}
                    \STATE return false;    \ENDIF
                \ENDFOR
                \STATE return true;
     \ELSE
        \STATE return false;
     \ENDIF
\end{algorithmic}
\end{algorithm}

\begin{algorithm}[h!]
\caption{: Function Execution} \label{algo_Execute}
\begin{algorithmic}
\STATE $void \ Execution( pid\_t \ k)$
    \STATE $trace\_t \ O$;
    \STATE $int \ l, i, j$;
    \COMMENT {Initialize all copies of the text}
    \IF {($InitOk==0$)}
            \FOR {\(( \ i=0; \ i < NbSites; \ i++)\)}
                \FOR {\((j=0; \ j<L; \ j++)\)}
                    \STATE   $text[i][j]=-1$;
                \ENDFOR
            \ENDFOR
              \STATE $InitOk=1$;
    \ENDIF\\
    \COMMENT{get the identifier of the $V[pid][k]$ (th) operation of
    $k$}
     \IF {$(pid != k)$}
               \FOR { \(( \ i=0, \ j=0; \ i < MaxIter \ \&\& j <= V[pid][k]; \ i++)\)}
                    \IF{ ($Operations[i].Owner \ == \ k$) }
                       \STATE $j++$;
                    \ENDIF
                \ENDFOR
                % \COM $i-1$ is the  identifier of the operation
                \STATE $O.numOp = i-1$;
                \STATE $O.posC = Operations[i-1].ipos$;
        \ELSE   \STATE $O.numOp = ns-1$;
                \STATE $O.posC = Operations[ns-1].ipos$;
                \FOR{$(i=0; \ i<NbSites; \ i++)$}
                    \STATE $Operations[ns-1].V[i]= V[pid][i]$;
                \ENDFOR
        \ENDIF
        \STATE $Transformation (O, List[pid], text[pid])$;
        \STATE $V[pid][k]++$;
\end{algorithmic} %\\

\end{algorithm}

\begin{algorithm}[h!]
\caption{: Function Transformation} \label{algo_Transformation}
\begin{algorithmic}
 \STATE $void \ Transformation \ (trace\_t \ \& \ op, trace\_t \ \& \ List[MaxIter], int[-1,1] \ \& \ t[L]) $
 \STATE $int \  i, len$;
 \FOR {($i=0,len=0;\ i<NbSites; \ i++$)}
      \STATE $len= len+V[pid][i]$;
 \ENDFOR
    \IF{($ len>0 \ \&\& \ pid != Operations[op.numOp].Owner$)}
            \STATE $TransformR \ (op, List, len)$;
    \ENDIF
    \STATE $Operation(op, List, len, t);$
\end{algorithmic}
\end{algorithm}

\begin{algorithm}[h!]
\caption{: Function Operation} \label{algo_Operation}
\begin{algorithmic}
 \STATE $void \ Operation(trace\_t \ \& \ op, \ trace\_t \ \& \
List[MaxIter], \ int \ len, \ int [-1,1] \ \& \ t[L])$
    \STATE $ int \ i;$
     \IF{ \((op.posC>=0 \  \&\& \ op.posC<L)\) }
        \IF { \((Operations[op.numOp].opr==Ins)\)}
        \FOR { \((i=L-1; \ i>op.posC; \  i--)\)} \STATE $t[i] = t[i-1]$;
        \ENDFOR
          \STATE  $t[op.posC]= Operations[op.numOp].x$;
         \ELSE
          \FOR { \((int \ i=op.posC; \ i <L-1; \ i++)\)}
                \STATE $t[i] = t[i+1]$ ;
          \ENDFOR
            \STATE    $t[L-1] = -1$;
         \ENDIF
     \ENDIF
    \STATE $List[len]=op$;
\end{algorithmic}%\\
\end{algorithm}

\begin{algorithm}[h!]
\caption{: Function TransformR} \label{algo_TransformR}
\begin{algorithmic}
 \STATE $void \ TransformR \ (trace\_t \ \& \ op, trace\_t \ \&  \ List[MaxIter],
 int \ len)$
 \STATE  $trace\_t \ List1[MaxIter]$;
 \STATE $int \  i$;
 \STATE  $bool \ Swap=false$;
 \STATE $ReOrder(op, List, len, List1, Swap)$;
 \IF{$(Swap)$}
     \FOR {($i=0; i<len; i++$)}
                \STATE $TransformR(List1[i], List1,i)$
     \ENDFOR
 \ENDIF
 \FOR {($i=0; i<len; i++$)}
 \IF {$(Concurrent(op,List1[i]))$}
  \IF {($algo == Ellis$)}
            \STATE $TEllis(op,List1[i])$;
  \ENDIF
  \IF {$(algo == Ressel)$}
                \STATE $TRessel(op,List1[i])$;
  \ENDIF
  \IF {$(algo==Sun)$}
                \STATE $TSun(op,List1[i])$;
  \ENDIF
  \IF {$(algo==Imine)$}
               \STATE $TImine(op,List1[i])$;
  \ENDIF
  \IF {$(algo == Suleiman)$}
    \STATE $TSuleiman(op,List1[i])$;
  \ENDIF
  \ENDIF
  \ENDFOR
\end{algorithmic}
\end{algorithm}

\begin{algorithm}[h!]
\caption{: Function ReOrder} \label{algo_ReOrder}
\begin{algorithmic}
\STATE $void \ ReOrder (trace\_t \ \& op, trace\_t \
List[MaxIter], int \ len, trace\_t \ \& \ List1[MaxIter], bool \
\& \ Swap)$ \STATE $int \ i, j=0, k=0$; \STATE $trace\_t \
List2[MaxIter]$; \STATE $Swap = false$;
 \COMMENT{put dependent operations on the top of the List1}
 \FOR {$(i=0; \ i<len; \ i++)$}
            \IF {($ not \ Concurrent(op,List[i])$ }
                \STATE  $List1[j].numOp= List[i].numOp$;
                \STATE  $List1[j].posC = Operations[List[i].numOp].ipos$;
                \IF {$(i\ !=\ j)$}
                  \STATE  $Swap = true;$
                \ENDIF
                \STATE  $j++$;
            \ELSE
                \STATE $List2[k].numOp= List[i].numOp$;
                \STATE $List2[k].posC = Operations[List[i].numOp].ipos$;
                \STATE $k++$;
            \ENDIF
 \ENDFOR \\
 \COMMENT {add list List2 at the end of List1}
 \FOR {($i=j; \ i<len; \ i++$)}
              \STATE   $List1[i].numOp = List2[i-j].numOp$;
              \STATE $List1[i].posC = List2[i-j].posC$;
 \ENDFOR
\end{algorithmic}
\end{algorithm}

\begin{algorithm}[h!]
\caption{: Function Concurrent} \label{algo_Concurrent}
\begin{algorithmic}
 \STATE $bool \ Concurrent(trace\_t \ op1, \ trace\_t \ op2)$
    \IF {($(Operations[op1.numOp].V[Operations[op1.numOp].Owner]
                    \geq $ \\
               \ \  $Operations[op2.numOp].V[Operations[op1.numOp].Owner])
                \ \& \&$ \\
                $(Operations[op2.numOp].V[Operations[op2.numOp].Owner]
                \geq$ \\
                \ \ $Operations[op1.numOp].V[Operations[op2.numOp].Owner])$)}
                \STATE $return \ true$
    \ELSE
    \STATE $return \ false$
    \ENDIF
\end{algorithmic}%\\
\end{algorithm}

\begin{algorithm}[h!]
\caption{: Function SymbolicExecution}
\label{algo_SymbolicExecution}
\begin{algorithmic}

\STATE $void \ SymbolicExecution( pid\_t \ k)$
    \STATE $int \ O, i, j$;\\
     \COMMENT {get the identifier of the $V[pid][k]$ (th) operation of
     $k$}
    \STATE \IF {$(pid != k)$}
                \FOR { ($ \ i=0, \ j=0; \ i < MaxIter \ \&\& \ j <= V[pid][k]; \ i++$)}
                    \IF{ ($Operations[i].Owner \ == \ k$) }
                       \STATE $j++$;
                    \ENDIF
                \ENDFOR \\
                 \COMMENT  {$i-1$ is the  identifier of the
                 operation}
                \STATE $O = i-1$ \;
        \ELSE   \STATE $Operations[ns].Owner=pid;$
                \FOR{$(j=0; j<NbSites; j++)$}
                    \STATE $ Operations[ns].V[j]= V[pid][j]$;
                \ENDFOR
                \STATE $O=ns;$
                \STATE $ns++;$
        \ENDIF
        \FOR {$(j=0,i=0; j<NbSites; j++)$}
            \STATE $i=i+V[pid][j];$
         \ENDFOR
          \STATE $Trace[pid][i]=O$;
    \STATE $V[pid][k]++$;
\end{algorithmic}
\end{algorithm}

%\begin{algorithm}
%\caption{Function SymbolicGuard}
%\begin{algorithmic}
%\STATE $bool \ garde(pid\_t \ k)$ \COMMENT {condition
%d'ex\'ecution d'une operation } \STATE $int O, i,j$;
%
% \IF {$(pid==k)$}
%    \STATE $return \ V[pid][pid] < Iter[pid]$;
%\ENDIF \COMMENT {rechercher la V[pid][k] i\`eme operation de k
%dans Operations}
% \IF {$(V[pid][k] < V[k][k])$}
%          \STATE  $j=0$;
%            \FOR {$(i=0; i< MaxIter && j<=V[pid][k]; i++)$}
%                \IF {$(Operations[i].Owner==k)$}
%                    \STATE $j=j+1$;
%                \ENDIF
%            \ENDFOR
%            \COMMENT {i-1 est le num\'ero de cette op\'eration}
%            \FOR { $(j=0; j<NbSites \ \&\& \ V[pid][j] >= Operations[i-1].V[j]; j++)$} ;
%            \ENDFOR
%         \IF {($j ==NbSites$)} \STATE $return \ true$;
%         \ENDIF
% \ENDIF
%    \STATE $return \ false$;
%\end{algorithmic}
%\end{algorithm}
%

\begin{algorithm}[h!]
\caption{: Function SymbolicExecution2}
\begin{algorithmic}
\STATE $void \ SymbolicExecution2(pid\_t \ k)$
\STATE $int \ n, i, j=0$;\\
 \IF { $(pid !=k)$}

    \FOR {$(i=0; i< MaxIter \ \&\& \ j<=V[pid][k]; i++)$}
            \IF {$(Operations[i].Owner==k)$}
                \STATE $j=j+1$;
            \ENDIF
    \ENDFOR

    \STATE $n=i-1$;
 \ELSE
    \STATE $n=ns$;
    \STATE $Operations[ns].Owner=pid$;
    \FOR {$(i=0; i<NbSites;i++)$}
            \STATE $Operations[ns].V[i]=V[pid][i]$;
    \ENDFOR
     \STATE $ns++$;
\ENDIF
 \FOR {$(i=0,j=0; i<NbSites; i++)$}
    \STATE $j=j+V[pid][i]$;
 \ENDFOR
 \STATE  $Trace[pid][j]=n$;
  \STATE  $V[pid][k]++$;

    \IF {$(j==MaxIter-1)$)}
        \FOR {$(i=0; i<NbSites \ \&\& \ not \ Stop; i++)$}
            \IF {$(i \ != pid)$}
                %\FOR{$(j=0; j<NbSites \ \&\& \ V[i][j]==Iter[j]; j++)$}
                %\ENDFOR
                \IF {$(j==NbSites)$}
                        \STATE $Stop=true$;
                \ENDIF
            \ENDIF
         \ENDFOR
    \ENDIF

\end{algorithmic}
\end{algorithm}
\end{document}